\documentclass[twocolumn]{IEEEtran}
\IEEEoverridecommandlockouts                              

\input required_files/setup.tex
  \ifx\LabelFigloaded\MYundefined\relax
  \else
   \fi

  \def\LabelFigloaded{\relax}

  \chardef\LabelFigCatAt\the\catcode`\@
  \catcode`\@=11

 \let\LabelFigwlog@ld\wlog
 \def\wlog#1{\relax}

 \ifx\\\MYundefined@
    \let\\\relax
 \fi

  \def\ms@g{\immediate\write16}

 \def\N@wif{\csname newif\endcsname }
 \def\Temp@ {\N@wif\ifIN@}
 \ifx\INN@\MYundefined@
    \else \let\Temp@\relax
 \fi
 \Temp@

  \def\IN@{\expandafter\INN@\expandafter}
  \long\def\INN@0#1@#2@{\long\def\NI@##1#1##2##3\ENDNI@
    {\ifx\m@rker##2\IN@false\else\IN@true\fi}%
     \expandafter\NI@#2@@#1\m@rker\ENDNI@}
  \def\m@rker{\m@@rker}
 
  \newtoks\Initialtoks@  \newtoks\Terminaltoks@
  \def\SPLIT@{\expandafter\SPLITT@\expandafter}
  \def\SPLITT@0#1@#2@{\def\TTILPS@##1#1##2@{%
     \Initialtoks@{##1}\Terminaltoks@{##2}}\expandafter\TTILPS@#2@}

 \def\Shifted@@#1#2#3{\setbox0=\hbox{#3}%
   \raise -\dp0\vbox {\kern-#2%
       \hbox {\kern#1\unhbox0\kern-#1}%
           \kern#2}}

 \newcount\gridcount
 \newbox\auxGridbox@ \newbox\hGridbox@ \newbox\vGridbox@
 \newbox\Labelbox@ \newbox\auxLabelbox@
 \newbox\Coordinatebox@
 \newtoks\Labeltoks@
 \newdimen\Wdd@ \newdimen\Htt@
 \newdimen\Wddd@ \newdimen\Httt@
 
 \def\Wr@{\immediate\write16}

 \newdimen\GL@wd
 \def\GridLineWidth#1{\GL@wd=#1}

 \def\gobble#1{}
 \def\EdgeErr@{\Wr@{}%
      \Wr@{\string\Edges\space argument
      1, 10, 100 or 1000 please\string!}%
      }

 \newcount\Edgect@

 \def\Sweepup#1\endSweepup{}

 \def\SetEdges@{%
    \edef\Zr@@s{\expandafter\gobble\number\Edgect@\empty}%
        \count255=0\Zr@@s\relax
        \ifnum\count255=\z@\else\EdgeErr@\show\tailtest\fi
        \count255=1\Zr@@s\relax
        \ifnum\count255=\Edgect@\relax\else\EdgeErr@\show\leadtest\fi
    \EdgGl@b\edef\Zr@s{\expandafter\gobble\Zr@@s\empty}
    \ifnum\Edgect@>\@ne\relax\EdgGl@b\let\L@Dc\empty
        \else\EdgGl@b\edef\L@Dc{\string.}\fi
    \ifnum\Edgect@>\@ne\relax
        \EdgGl@b\edef\Edgescale@##1{\divide##1 by \Edgect@}%
        \else\EdgGl@b\edef\Edgescale@##1{}\fi
    }

 \def\Edges#1{\Edgect@=#1\relax
     \let\EdgGl@b\global \SetEdges@}

 \Edges{1}

 \def\hhrule{\hrule height \GL@wd\vskip-.\GL@wd}

 \def\hRule@{%
   \advance\gridcount -2%
   \vfil\hhrule\vfil
   \llap{\smash{\raise -2.5pt
     \hbox{\L@Dc\number\gridcount\Zr@s\kern2pt}}}%
   \hhrule
   }

\def\vvrule{\vrule width \GL@wd \kern-\GL@wd}

 \def\vRule@{\advance\gridcount 2%
   \hfil\vvrule\hfil
   \setbox\auxGridbox@=\vbox to 0pt
      {\vskip \Htt@\vskip 2pt
        \hbox to 0pt{\hss\L@Dc\number\gridcount\Zr@s\hss}\vss}%
      \wd\auxGridbox@=0pt \box\auxGridbox@
   \vvrule
   }

 \def\PlaceGrid@@{\gridcount=10 
  \setbox\hGridbox@=\hbox{%
        \hbox{%
             \hskip-.4pt\vrule
             \vbox to \Htt@{%
               \offinterlineskip\parindent=\z@\relax
               \hbox to \Wdd@{\hfil}
               \hRule@\hRule@\hRule@\hRule@
               \vfil\hhrule\vfil}%
             \vrule\hskip-.4pt}
    }%
  \gridcount=0%
  \setbox\vGridbox@=\hbox{%
      \vbox{\offinterlineskip\parindent=0pt\hsize=0pt
         \vskip-.4pt\hrule%
         \hbox to \Wdd@{%
                 \vtop to \Htt@{\vfil}%
                 \vRule@\vRule@\vRule@\vRule@
                 \hfil\vvrule\hfil}%
         \hrule\vskip-.4pt}}%
  \wd\hGridbox@=0pt\ht\hGridbox@=0pt
  \wd\vGridbox@=0pt\ht\vGridbox@=0pt
  \hbox{\box\hGridbox@\box\vGridbox@}%
  }

 \def\LabelsGlobal{\def\LabGl@b{\global}}
 \def\LabelsLocal{\def\LabGl@b{}}
 \LabelsGlobal 

 \def\SetLabels#1\endSetLabels{%
   \LabGl@b\Labeltoks@={#1()\\}%
   }

 \LabGl@b\Labeltoks@={()\\}

 \def\ShowGrid{\LabGl@b\let\PlaceGrid@\PlaceGrid@@}
 \def\HideGrid{\LabGl@b\let\PlaceGrid@\relax}
 \def\Grids{\ShowGrid\LabGl@b\let\GridSwitch@\ShowGrid}
 \def\noGrids{\HideGrid\LabGl@b\let\GridSwitch@\HideGrid}

 \noGrids

 \def\bAdjust@@{%
     \setbox\auxLabelbox@=\hbox{\raise \dp\auxLabelbox@
            \box\auxLabelbox@}}
 \def\bAdjust@{\let\vAdjust@\bAdjust@@}

 \def\eAdjust@@{\dimen0=-.5\ht\auxLabelbox@
     \advance\dimen0 by .5\dp\auxLabelbox@
     \setbox\auxLabelbox@=
            \hbox{\raise\dimen0\box\auxLabelbox@}}
 \def\eAdjust@{\let\vAdjust@\eAdjust@@}

 \def\tAdjust@@{%
     \setbox\auxLabelbox@=\hbox{\raise-\ht\auxLabelbox@
            \box\auxLabelbox@}}
 \def\tAdjust@{\let\vAdjust@\tAdjust@@}

 \let\vAdjust@\relax

 \def\lAdjust@{\let\hAdjust@\rlap}
 \def\rAdjust@{\let\hAdjust@\llap}

 \let\hAdjust@\relax\let\vAdjust@\relax

 \def\FetchLabel@#1(#2)#3\\{%
     \IN@0#2@@\ifIN@
        \setbox0=\hbox{\ignorespaces#1#3\unskip}%
        \ifdim\wd0>0pt
           \ms@g{}%
           \ms@g{ !!! Bad label(s)? !!!}%
           \message{ #1(#2)#3}%
        \fi
        \def\LabelMole@##1\endFetchLabel@{%
            \IN@0()\\@##1@%
            \ifIN@\def\Temp@{\FetchLabel@##1\endFetchLabel@}%
            \else\def\Temp@{}%
            \fi
            \Temp@
           }%
     \else
       \ignorespaces#1\unskip
       \setbox\auxLabelbox@=%
         \hbox to 0pt{\hss\ignorespaces\hAdjust@
          {\ignorespaces#3\unskip}\hss}%
       \vAdjust@
       \let\hAdjust@\relax\let\vAdjust@\relax
       \AugmentLabelBox@@{#2}%
       \ht\Labelbox@=0pt\dp\Labelbox@=0pt
       \let\LabelMole@\FetchLabel@%
     \fi\LabelMole@}

 \newtoks\XYSep@ 
 \def\SetXYSeparator#1{%
     \IN@0#1@@\ifIN@\XYSep@{*}%
     \else
     \XYSep@{#1}%
     \fi
     }

 \SetXYSeparator*

 \def\AugmentLabelBox@@#1{%
     \IN@0\the\XYSep@ @#1@\ifIN@
       \SPLIT@0\the\XYSep@ @#1@%
       \setbox\Labelbox@=\hbox to 0pt{%
         \unhbox\Labelbox@
         \Shifted@@{\the\Initialtoks@\Wddd@}%
         {\the\Terminaltoks@\Httt@}%
         {\box\auxLabelbox@}}%
     \else
         \ms@g{}%
         \ms@g{ !!! Bad insertion point. !!!}%
         \message{ (#1\ this point was rejected.)}%
     \fi
    }

 \def\FetchOption@#1[#2]#3\endFetchOption@{%
    \def\temp{#1}
    \ifx\temp\empty
       \Edgect@=#2\relax
       \let\EdgGl@b\relax
       \SetEdges@
       \Cleaner@#3%
    \fi}

 \def\Cleaner@#1[@]{\Labeltoks@{#1}}
     
 \def\PlaceLabels@@{\mathsurround=0pt
     \def\Cr@{\\}%
     \let\L\lAdjust@\let\R\rAdjust@
     \let\B\bAdjust@\let\E\eAdjust@\let\T\tAdjust@
     \expandafter\FetchOption@\the\Labeltoks@[@]\endFetchOption@
     \Wddd@=\Wdd@ \Edgescale@\Wddd@ 
     \Httt@=\Htt@ \Edgescale@\Httt@
     \expandafter\FetchLabel@\the\Labeltoks@\endFetchLabel@
     \box\Labelbox@
     }%

 \let \PlaceLabels@\PlaceLabels@@

 \def\AffixLabels#1{\setbox\Coordinatebox@=\hbox{#1}%
      \Wdd@=\wd\Coordinatebox@ \Htt@=\ht\Coordinatebox@
      \advance\Htt@ \dp\Coordinatebox@
      \hbox{\copy\Coordinatebox@\kern-\Wdd@ 
           \Shifted@@{0pt}{-\dp\Coordinatebox@}%
           {\PlaceLabels@\PlaceGrid@}%
           \kern\Wdd@}%
      \GridSwitch@ 
      \LabGl@b\Labeltoks@{()\\}%
      }
 
   \let\wlog\LabelFigwlog@ld   
   \catcode`\@=\LabelFigCatAt  

                                By

              Raymond S\'eroul <A18645@FRCCSC21.BITNET>
                                and 
              Laurent Siebenmann <lcs@topo.math.u-psud.fr>
    
              VERSIONS: July 1991, Oct 1991, Jan 1992, July 1992

INTRODUCTION

      This labelling package is intended for TeX users who
rely on non-TeX sources for for their graphics inserts.  It
provides means for adding TeX labels to such inserts with a
minimum of fuss. 

       For most labels, TeX users have in the past found it
reasonably convenient to rely on non-TeX sources. Typical
occasions when an inescapable need for TeX labels seemed to
arise are

 (a) when the graphics program lacks certain exotic or complex
mathematical symbols

 (b) when the very highest typographical quality is wanted for the
labels

 (c) when labels included with the graphics fail to print, 
 and you cannot figure out why (cf. boxedeps.doc).  The labels
 provided by labelfig.tex are 100

       Since this package first appeared, many users, who in the
past scarcely dreamed of using TeX labels, have come to use
nothing but.  So it is now appropriate to add

Intoxication Warning:  TeX labels may be addictive and expensive. 

     If you have a fast preview you may disagree, and even find
that this package provides an agreeable paste-up environment; see
extra applications at end.

     Note to publishers: It is possible and convenient to ultimately
export the TeX labels produced by labelfig.tex to become an integral
part of the EPS file. This is often desired by a publisher who typically
uses an "upmarket" graphics or page layout program, with which the
staff is skilled in perfecting figures.  See Appendix I for
a recipe.

     The authors are grateful to Patrick Ion of Math Reviews for
helpful comments and encouragement.

BASIC INSTRUCTIONS

    After reading in the macro file using

preview or proof your figure with a coordinate grid printed on
top, by typing the following:

    \ShowGrid  
    \AffixLabels{<the graphics insertion>}

Here <the graphics insertion> is what you would type to insert
the graphics object alone without the grid.  This must provide
for the space around it. For example <the graphics insertion>
might well be \BoxedEPSF{MyFigure scaled 700} using the
boxedeps.tex macro package (from same source); this provides a
TeX box containing the encapsulated PostScript insert specified by
the file MyFigure. \AffixLabels{...} provides the grid (supposing
\ShowGrid is present) and later, once you have specified labels
using the grid, it will "tack on" the labels.

     The grid is a sort of (usually elongated) checkerboard of
ten rows and ten columns and its (internal) partitions are by
default numbered  .1, ... ,.9  both horizontally (X-coordinate
running left to right) and vertically (Y-coordinate running bottom
to top).  Thus the points enclosed by the grid correspond to the
points of the unit square in the cartesian "X-Y" plane, the lower
left corner corresponding to the origin (0,0).  By extrapolation,
the full page corresponds to a larger rectangle in the plane.

     These coordinates serve to position labels as follows.
Before the \AffixLabels{...} command type label specifications:

  \SetLabels
   (<X-coordinate>*<Y-coordinate>) <first label> \\
   .
   .
   .
   (<X-coordinate>*<Y-coordinate>)  <last label> \\
  \endSetLabels

Each row specifies one label and is terminated by \\.  In each
row, the position indicator comes first; it is written as a
standard cartesian point except that the X- and Y- coordinates
are separated by * rather than a comma because TeX allows a
comma as decimal point. There are no dimension units to specify
as the unit is the grid itself.

     By default, this cartesian point specifies where the middle
of the baseline of the label will be located.  However if you precede
the point by \L [or \R] the left [or right] edge of the baseline will
be located there. Similarly you may also precede the point by \T, \E,
or \B to vertically align the top equator or bottom of the label box
at the specified point.  This gives nine standard positions of
the label with respect to the insertion point --- corresponding to
the eight principle points of the compas and the center

                     \L\T     \T      \R\T

                     \L\E     \E      \R\E

                     \L\B     \B      \R\B

But this neglects the default "baseline" level of TeX,
giving potentially three more positions

                     \L    <no tag>   \R

For text, the baseline level is often the preferred. Its relation to
the others is variable. It will often coincide with the bottom level,
as happens for "X".  But it is often distinct, as for "g", in which
case you have in all 12 distinct positions rather than 9.

     It is convenient to think of this specification of label
position as attaching the label by a thumb-tack to the coordinate
grid. There are up to twelve positions of the thumb-tack on the
label, while the position of the thumb-tack on the coordinate grid is
arbitrary.  Normally, one choses the position of the thumb-tack on
the label to be the one that is the closest to the item being
labeled.  There are good reasons for this "rule of thumb":

   (a)  It facilitates correct positioning at first try.

   (b)  If the scale of the figure must be altered after labels
have been affixed, the labels have a good chance of remaining well
positioned.

   (c)  The visible grid need not extend beyond the "bounding box"
for the figure, because the best preferred position is always
(at least almost) within the bounding box .

The second reason is particularly important. Indeed it often
happens that scale has to be altered after labelling begins, in
order to either provide space for the labels, or to adjust
proportions between the labels and the figure.  (The size of labels
is unaffected by scaling.)

     Here is an artificial but self-contained test which uses
TeX rules to make a graphics object.

TEST

    Do not skip this!

 \def\FrameIt#1{\hbox{\vrule$\vcenter {\hrule\kern3pt%
             \hbox {\kern3pt #1\kern3pt}%
               \kern3pt\hrule}$\relax\vrule}}

 \def\Caption#1#2{\FrameIt{%
       \vtop {\hsize=#1\relax \parindent=0pt
         \leftskip=0pt \rightskip=0pt plus15pt
         \parfillskip=0pt
         \lineskip=1pt\baselineskip=0pt
         #2}}}

 \def\FirstQuadrant{\hbox to 100pt{\vrule\vbox to 100pt{%
        \hbox to 100pt{\hfil}\vfil\hrule}\hss}}

  \SetLabels
    \R(.5*.2) $\zeta\,\cdot$\\
    (.9*-.10) $\xi$\\
    \R(-.03*.9) $\eta$\\
    \T(.5*.9) \Caption{70pt}{%
          \it The norm of
          $g(\xi+i\eta)$ is indicated on
          contours of this invisible surface.}\\
  \endSetLabels

  \AffixLabels{\FirstQuadrant}

  \end

  Note that the coordinates to use for labels are indicated on the
edges of the grid (when visible) corresponding to the conventional
x- and y- axes of the Cartesian plane. By default the grid is
1-by-1. However, by the command \Edges{100}, you can change this
to 100-by-100 and many users find this alternative most
convenient. Place the command \Edges{...} in your style file (or
header) since its effect is is global. Other possible edge values
are 10 and 1000.

  If you use the command \Edges{...} at all, do so with care.  For
if you accidentally delete an \Edges{...} command your labels will
abruptly be badly misplaced and may logically but mysteriously
generate "dimension too big" errors under TeX and "off page" errors
under your driver.  

  You can dictate the edgescale for an individual figure by giving
the scale in brackets immediately after \SetLabels.  Thus, to
import into an article using say \Edge{100} a figure labelled using
another edgescale, say the original 1-by-1 default, you can use
\SetLabels[1]...\endSetLabels.

GETTING IT DOWN PAT

     Complicated labeling deserves the same respect as
complicated mathematics.  Do not expect it to come out perfect the
first time!  What is needed in either case is a mechanism to
repeatedly typeset troublesome pieces.

     One mechanism is always available.  One does complicated
labelling in a separate "test" file involving just the figure being
labelled;  a texpert will know how to \dump TeX's current state as
a temporary format that restarts rapidly at each retry.  Usually,
one then pastes the completed labelled figure back into the main
TeX file, but, of course, one can also \input it as an auxiliary
file.

     If you do not have a TeXpert at handy, here is a first
approximation to an efficient setup. By deletions reduce a copy
of your article to just a few lines before and after the figure.
Now label the figure, and finally, copy and paste the labelled
figure to the original article. Then copy the next figure to label
into this testbed and repeat. The TeXpert can improve the  speed
at which TeX starts up, by compiling a format specifically for
your article; just one caution: best NOT include in the format
ephemeral details of setup like \Set<mydriver>ArtSpecials (from
boxedeps.tex because this reads  figure dimensions which you may
change during your work session.

     An improved mechanism to repeatedly typeset troublesome
pieces is now available on the Macintosh; it is called LinoTeX;
see the same ftp sources.  It could be set up on many types
of computer.

     Before using labelfig.tex to attach labels to a graphics
object inserted using boxedeps.tex or BoxedArt.tex, make it a
firm rule to carefully adjust the bounding box using the trimming
commands of these packages, and also at least tentatively scale
and position the object. Beware of changing the grid inadvertently
after the labels have been positioned.  For example, correcting
the bounding box of a PostScript graphics object can foul up the
labels by changing the coordinate grid to which the labels are
attached. This is particularly true for the trimming  commands of
boxedeps.tex and BoxedArt.tex. However, as noted already, change
of scale is much less disruptive, and modest adjustments should be
well tolerated.

     Sometimes the labels protrude so far from the bounding box
of a figure that the figure has to be repositioned.  Best do this
by ad hoc spacing, say using \hglue and \vglue; altering the
bounding box would create a vicious circle.

     Remember that you are responsible for preventing labels
from overlapping. You are responsible for all label typography
including size and style. A label is really just about anything
that can be put in a TeX box. Note that spaces at the beginning
and end of labels will normally be suppressed; if you really want
them you must protect them with TeX braces.

     This package temporarily sets the \mathsurround parameter
of TeX to zero  while the labels are being affixed. This is done
because nonzero \mathsurround space would influence the position
of left and right aligned labels; then, when a texpert or printer
modifies mathsurround, diagram labeling might be disastrously
altered. There is a small price to pay involving labels that are
formatted as caption boxes including mathematics: you  may want or
need to specify an explicit mathsurround space within the caption
box; it will not influence anything outside.

     Those hostile to the use of * as separator between
the X and Y coordinates of label insertion points, are free to
impose another using \SetXYSeparator{<the new separator>}.  
Americans may prefer "," to "*" since they never use a 
comma as a decimal point; on the other hand, * may be more visible.

APPENDIX (I)  MERGING labelfig.tex LABELS INTO AN EPSF GRAPHICS OBJECT.

     As promised in the introduction, here is a recipe useful for
publishers. It works at least on Macintosh and at least for vectorized
graphics and Adobe type1 fonts.  (There is surely a similar recipe for
PCs under MSWindows.)

 (a)  Use boxedeps.tex utility to integrate the figure given by the eps
file, "x.eps" say, with a visible frame around it.  See
\ShowDisplacementBoxes command in boxedeps.tex.  To get precise results
automatically it is important to use the \Trim... commands of
boxedeps.tex making the "DisplacementBox" neatly fit the figure.

 (b)  Use the TeX printer driver and LaserWriter (versions >= 8.1.1) to
export to an EPSF the DVI page containing the integrated, labelled
figure. You now have an EPS file  "xx.eps"  that contains too much, and at
the wrong scale, and at wrong position.

 (c)  Convert the EPSF to an Adode Illustrator format EPSF using
the shareware utility called epsConvert by Sam Weiss
1993-- (currently $25).

 (d)  In Illustrator (or a compatible program), group the labels and the
"DisplacementBox"; copy them to the clipboard and paste them into "x.ps".
This step requires that all the label fonts be "visible to the Macintosh.

 (e)  Translate and scale the pasted group consisting of the labels plus
the "DisplacementBox" so as to make the "DisplacementBox" the bounding
box of (labelless) figure represented by "x.eps".  At this point the
labels will be correctly placed on the figure "x.eps".

 (f)  Ungroup and delete the "DisplacementBox".  The result is the
desired single EPS file, "x+.eps" say, It contains the original figure
plus its labels.  

     Using grouping and ungrouping appropriately in "x+.eps", a
publisher's staff can very efficiently improve label positions etc.

APPENDIX II)  SOME EXOTIC APPLICATIONS

     The grid of labelfig.tex is analogous to a light-table in
classical page makeup with wax or latex glue.  In principle, you
can use it to compose any page from its indivisible parts.  This
even has some of the artisanal charm of classical paste-up
provided you have a fast screen preview to make the process
"interactive".

     In practice labelfig.tex is a tool for nonstandard jobs.
Here are a few going beyond the labelling already discussed.

(I)  GRAPHICS INTEGRATION.

     This is accomplished by treating the imported graphics
objects as labels.  The underlying graphics object is then
typically an empty  \vbox to <dimension>{\vfill} in a TeX
\midinsert...\endinsert construction.  A label line
might be of the form

   (.1*.1) \\

The exact form of the special command varies from driver to
driver.  However, in the case of encapsulated PostScript graphics
(EPSF norm), by relying on boxedeps.tex, one can have the
following standard syntax (independant of driver  (see
boxedeps.doc for details.
  
  (.1*.1) \BoxedEPSF{MyFigure scaled <scale in mils>}\\

This may be slow since it requires TeX to read the PostScript
file to read bounding box using many complex macros.  So you
may want to try

  (.1*.1) \EPSFSpecial{MyFigure}{<scale in mils>}\\

which is fast and driver independant, but it squashes the
bounding box, normally to its lower left corner.

     Similarly for graphics of the Macintosh PICT norm ---
using BoxedArt.tex (same sources) in place of boxedeps.tex.

     This approach to integration is to be recommended when
one is assembling a composite graphics object.

 (II)  COMMUTATIVE DIAGRAM ENHANCEMENT

     Commutative diagrams or arrays of mathematical objects
connected by arrows of various sorts are common in mathematics.
The mathematical objects require the use of TeX.  Recently TeX
acquired a good collection of arrows of all slopes --- that of
LamSTeX --- plus pwerful macros to build the diagrams.

     However, even the LamSTeX collection is often
inadequate; it lacks for example double shafted arrows, dotted
arrows and curved arrows. Fortunately it is possible to produce
such arrows on an individual basis using sophisticated graphics
programs such as Illustrator and AldusFreehand (both serving
the EPSF norm) or using Metafont (with its public domain norm).
Since the creation of each new arrow is a work of love, you
probably want to limit the number of arrows by using LamSTeX
for most arrows. The 40K commutative diagram module of LamSTeX
has been adapted to work with AmSTeX and a copy may be posted
with LabelFig and related files. Unfortunately no one has yet
offered a version that works with Plain TeX or LaTeX.

       Suffice it here to say that when the exotic arrow has
been somehow imported into TeX, labelfig.tex treats it as a
label that one affixes to the commutative diagram.  Two other
steps will be treated in separate notes, namely the matter of
extracting the dimension specifications for the arrow and the
construction of the arrow --- for these steps are far from
unique and often depend intimately on your computer environment. 
Notes for the Macintosh-Textures-Illustrator combination are
found in the file ExoticArrows.doc.

 (III) NESTING 

Ingenuity pays off in exploiting labelfig.tex. One can
mix graphics and typography quite freely.  labelfig.tex is good
for freeform or overlapping arrangements, while boxedeps.tex (or
BoxedArt.tex) is best for regimented non-overlapping
arrangements --- and the two can be combined.

     The default behavior of labelfig.tex is not ideal 
for nesting objects, because to prevent trouble for beginners
the register for labels is globally cleared when \AffixLabels
concludes.  But there are switches available

      \LabelsGlobal      \LabelsLocal

which change this.  To understand this, extend the above test 
by something like:

 \LabelsLocal

 \SetLabels
    (.5*.5) AAA\\
 \endSetLabels

 {
 \SetLabels
    (.5*.5) ZZZ\\
 \endSetLabels
   \AffixLabels{\FirstQuadrant}
 }

   \AffixLabels{\FirstQuadrant}

     There are however potential pitfalls.  Neither
labelfig.tex nor boxedeps.tex has been tested under extreme
conditions. Problems may occur if their procedures are
indiscriminately nested. For boxedeps.tex (not labelfig.tex)
there is a precise cause for worry, namely many of its
variables are "global", which means that TeX braces will not
provide the protection one might expect.

COMMAND SUMMARY FOR labelfig.tex

  Here [...] means optional (one or zero)
       [...]* means any number of such constructs

  \SetLabels
    [[<P>](<X><Sep><Y>) <label> \\]*
  \endSetLabels
  \ShowGrid  
  \AffixLabels{<the figure>}

   --- <P> is tack position, one of eleven or empty
              order irrelevant

                   \L\T      \T      \R\T

                   \L\E      \E      \R\E

                     \L               \R

                   \L\B      \B      \R\B

   --- (<X><Sep><Y>) insertion point;
  <Sep> is separator, = * by default;
  \SetXYSeparator{<Sep>} changes it.
   <X> and <Y> are real numbers

  --- <label> a label to attach 

  --- <the figure> the figure to label 

  \GlobalLabels (default)     
  \LocalLabels  setting for nested constructs.

 \Grids makes ALL grids appear; \HideGrid then makes just next disappear.
 \noGrids returns to default.  The commands are always global.

 \GridLineWidth{<dimension>} adjusts width of grid lines. Default is very
small, to give "hairline" effect. If your grid lines are missing try
setting \GridLineWidth{1pt}.

 \Edges#1 globally changes the edge size of all grids to the numerical 
value #1, which must be 1, 10, 100, or 1000.  The default is 1.

VERSION HISTORY.
 --- Jan 1993: \Edges#1 and [??] option after \SetLabels
 --- July 1992: \Grids, \noGrids, \HideGrid;
       Gridlines become hairlines; \GridLineWidth{<dimension>}.
 --- Oct 1991, Jan 1992: \SetXYSeparator{<Sep>},  \LabelsGlobal,
       \LabelsLocal.
 --- July 1991: first release

Address for bugs and other feedback:

        Raymond S\'eroul
        IREM and Lab. de Typographie Informatise
        Univ. Rene Descartes
        Strasbourg

    Tel 33-88-41-63-45
    Email:  A18645@FRCCSC21.BITNET

        Laurent Siebenmann
        Mathematique, Bat. 425,
        Univ de Paris-Sud,
        91405-Orsay,
        France

    Tel 33-1-6941-7949; 
    Email: lcs@topo.math.u-psud.fr

\usepackage[dvips]{graphics}
\usepackage[dvips]{graphicx}
\usepackage{latexsym}
\usepackage{graphics}
\usepackage{subfigure}
\usepackage{epsf}
\usepackage{epsfig}
\usepackage{amssymb,amsmath}
\usepackage{fancybox}
\usepackage{fancyhdr}
\usepackage{psfrag}

\def\scalefig#1{\epsfxsize #1\textwidth}
\def\calE{{\mathbb E}}
\def\calV{{\cal V}}

\def\Emsc{{\cal E}}
\def\calP{{\cal P}}
\def\calR{{\cal R}}

\def\calV{{\cal V}}
\def\calG{{\cal G}}

\def\Smsc{{\cal S}}
\def\calS{{\cal S}}
\def\calY{{\cal Y}}
\def\calZ{{\cal Z}}

\def\hcY{\hat{\cal Y}}
\def\bs{{\bf f}}
\def\bP{{\bf P}}
\def\lm{{\lambda}}
\def\Lm{{\bf \Lambda}}
\def\blm{{\bf \Lambda}}
\def\bflm{\overline{\lambda}}
\def\bs{{\bf s}}

\def\bT{{\bf T}}
\def\defeq{\stackrel{\triangle}{=}}
\def\bB{{\bf B}}

\def\bS{{\bf S}}

\def\hbS{\hat{\bf S}}

\def\bDel{\bar{\bf \Delta}}

\def\eps{{\epsilon}}
 
\newcommand{\ben}{\begin{enumerate}}
\newcommand{\een}{\end{enumerate}}
\newcommand{\beq}{\begin{equation}}
\newcommand{\eeq}{\end{equation}}
\newcommand{\nn}{\nonumber}
\newcommand{\beqa}{\begin{eqnarray}}
\newcommand{\eeqa}{\end{eqnarray}}
\newcommand{\beqq}{\begin{eqnarray*}}
\newcommand{\eeqq}{\end{eqnarray*}}
\newcommand{\bit}{\begin{itemize}}
\newcommand{\eit}{\end{itemize}}

\title{\bf Anonymous Networking amidst  Eavesdroppers}
\author{   Parvathinathan Venkitasubramaniam, Ting He and Lang
Tong \\
\it School of Electrical and Computer Engineering\\
\it Cornell University, Ithaca, NY 14853\\
\it Email : \{pv45, th255, lt35\}@cornell.edu
\thanks{{\scriptsize This work is supported in part  by the National
Science Foundation under awards CCF-0635070 and CCF-0728872, 
and the U. S. Army Research Laboratory under
the Collaborative Technology Alliance Program 
DAAD19-01-2-0011. Part of the results in this work were presented in at Allerton 2006 and Allerton 2007.} }
}

\begin{document}
%

\maketitle

\begin{abstract}
The problem of security against timing based traffic analysis in wireless networks is considered in this work. An analytical measure of anonymity in eavesdropped networks is proposed using the information theoretic concept of equivocation. For a physical layer with orthogonal transmitter directed signaling, scheduling and relaying techniques are designed to maximize achievable network performance for any given level of anonymity. The network performance is measured by the achievable relay rates from the sources to destinations under latency and medium access constraints. In particular, analytical results are presented for two scenarios:

For a two-hop network with maximum anonymity, achievable rate regions for a general $m\times 1$ relay are characterized when nodes generate independent Poisson transmission schedules. The rate regions are presented for both strict and average delay constraints on traffic flow through the relay. 

For a multihop network with an arbitrary anonymity requirement, the problem of maximizing the sum-rate of flows (network throughput) is considered. A selective independent scheduling strategy is designed for this purpose, and using the analytical results for the two-hop network, the achievable throughput is characterized as a function of the anonymity level. The throughput-anonymity relation for the proposed strategy is shown to be equivalent to an information theoretic rate-distortion function.
\end{abstract}

\IEEEpeerreviewmaketitle

\section{Introduction}
Traffic analysis attacks are carried out by eavesdroppers monitoring node transmissions to obtain networking information such as source-destination pairs and paths of data flow. Traffic analysis has played a prominent role in modern warfare \cite{West:Book} and its adverse effects on computer networks is well documented in literature \cite{Voydock&Kent:83ACM, Raymond:01, Sun&etal:02, Mathewson&Dingledine:04PET}. For example, the weaknesses of protocols for web browsing \cite{Sun&etal:02, Felten&Schneider:00CCS} and SSH \cite{Song&Wagner&Tian:01} have been exposed through traffic analysis. 

The primary focus of this work is an analytical approach to security against traffic analysis in wireless networks and the design of provably secure countermeasures. Owing to the unprotected medium of communication, eavesdropping node transmissions in wireless networks is easy and undetectable. Although cryptography can be used to prevent analysis based on contents or packet lengths (see Section \ref{sec:related}), the knowledge of transmission epochs alone can reveal critical information such as paths of information flow. We address the problem of designing {\it anonymous} transmission schedules and relaying strategies to counter the transmission epoch based inference of data flows by eavesdroppers. 

\begin{figure}[h]
\centering
\subfigure[Sources $S_1,S_2$ transmit packets to Destinations $D_1,D_2$ through Relay $B$]{
\begin{psfrags} 
\psfrag{a1}[c]{  $S_1$}
\psfrag{a2}[c]{  $S_2$}
\psfrag{b}[c]{  $B$}
\psfrag{ya1}[c]{  $$}
\psfrag{yb1}[c]{  $$}
\psfrag{yan}[c]{  $$}
\psfrag{ybn}[c]{  $$}
\psfrag{yc1}[c]{  $$}
\psfrag{ycn}[c]{  $$}
\psfrag{d1}[c]{  $D_1$}
\psfrag{d2}[c]{  $D_2$}
\scalefig{.28450920}\epsfbox{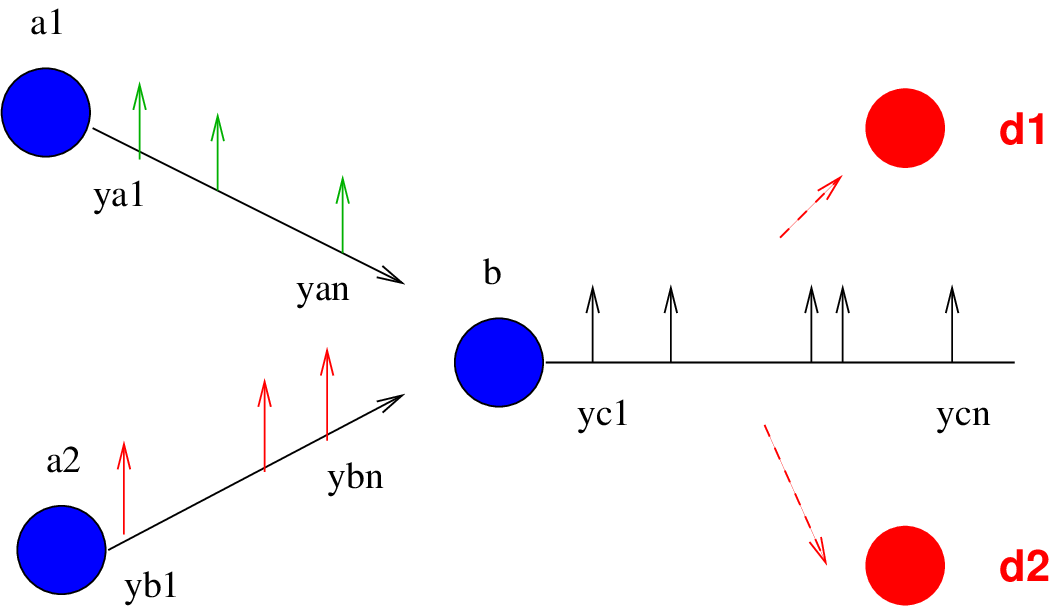} 
\end{psfrags} }\\
\vspace{.2em}
\subfigure[Achievable Rate Region : The horizontal and vertical boundaries are due to rate constraints $C_1,C_2$ on nodes $S_1,S_2$. The sum-rate constraint is due to stability at relay $B$. The inner region (in dotted line) represents the achievable rate region with independent scheduling that we wish to characterize.]{
\begin{psfrags}
\psfrag{ra}[c]{  $R_1$}
\psfrag{rb}[c]{  $R_2$}
\psfrag{a1r}[c]{  $R_1\leq C_1$}
\psfrag{a2r}[l]{  $R_2\leq C_2$}
\psfrag{sumr}[l]{  $R_1+R_2\leq C_B$}
\psfrag{sec}[c]{  \large ?}
\scalefig{.3283245}\epsfbox{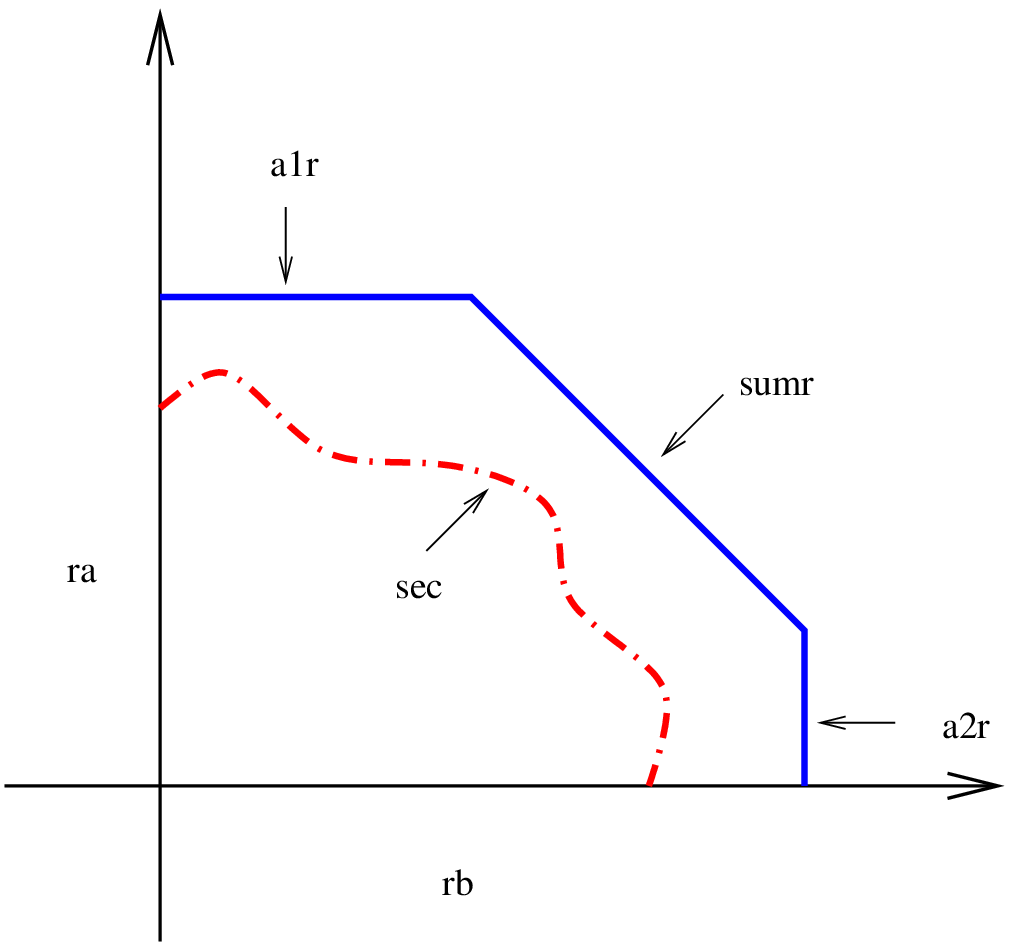}
\end{psfrags}}
\caption{Two Hop Relay Network \label{fig:twohop_example}}
\end{figure}

The challenge in designing {\it anonymous} transmission strategies is to adhere to the networking constraints while hiding information from eavesdroppers. Wireless networks are subject to constraints on medium access, latency and stability, which generally result in a high correlation across transmission schedules of nodes in a path. The need for anonymity however necessitates that paths are not revealed by correlation of transmission schedules. These contrasting paradigms result in a tradeoff between anonymity and network performance. For example, consider the simple two hop setup shown in Fig. \ref{fig:twohop_example}, wherein node $B$ relays packets received from nodes $S_1$ and $S_2$ subject to a strict delay constraint. Assuming the nodes use orthogonal channels, if the transmission rates $R_1,R_2$ are bounded, then the rates of packets that can be relayed successfully is given by a pentagon (solid line in Fig. \ref{fig:twohop_example}). Rates in this region are achieved if the relay transmits every received packet after a small processing delay. It is easy to see that such a strategy would result in a high correlation between the source and relay schedules. If, in addition to the networking constraints, the source and relay schedules are forced to be statistically independent, an eavesdropper would not detect correlation across schedules, thus hiding the relaying operation. The delay constraint may, however, result in packet drops or require dummy transmissions thereby reducing the achievable relay rates.

The relaying operation of Figure \ref{fig:twohop_example} represents the basic component in wireless networking, and the characterization of the achievable rate region with provable anonymity is one of the contributions of this work. The example highlights that providing anonymity in communication requires a reduction in communication rates. A primary goal of this work is to characterize this trade-off between anonymity and network performance. An analytical approach for the characterization requires a quantifiable notion of anonymity, which we measure using the uncertainty in networking information (active routes in the network) inferable by the adversary. The example discussed suggests a simple technique to provide perfect anonymity by letting all nodes generate statistically independent schedules, but this strategy may not provide scalable performance for large networks. Our goal is to design transmission strategies that sacrifice minimum performance while maintaining a certain level of anonymity.

\subsection{Main Contributions}\label{sec:results}
We propose an analytical framework for anonymous scheduling against traffic analysis in wireless networks. In particular, we define a mathematical notion for anonymity of routes, based on Shannon's equivocation \cite{Shannon:49BSTJ}, when eavesdroppers observe transmission epochs of all nodes in the network. The main results obtained under this model are divided into two segments.

Assuming maximum anonymity requirement, we design scheduling and relaying strategies for a two hop multiple source single relay system (see Fig. \ref{fig:two_hop}) when nodes use orthogonal transmitter directed signaling.  In particular, when the transmission schedules of nodes are independent Poisson processes, we characterize the achievable rate region analytically. Although independent Poisson scheduling may not be optimal for a strict delay constraint on the relay, we show that, under certain physical layer conditions, the achievable relay rates are optimal for an average delay constraint.

For a general multihop network, we propose a randomized scheduling strategy for any given level of anonymity $\alpha$, and utilizing the results of the two hop system, characterize the achievable sum-rate of data flows as a function of $\alpha$. Our key result in this framework shows the equivalence between the sum-rate anonymity tradeoff and information theoretic rate-distortion. 

The connection between rate distortion and anonymous networking is not tied to our strategy and can be explained using a general intuition. The objective of the rate-distortion problem is to generate fewest number of codewords for a set of source sequences, such that the corresponding reconstruction sequences satisfy a specified distortion constraint. The idea is to divide the set of source sequences into fewest number of bins such that the distortion between each sequence in a bin and the reconstruction sequence is less than the specified constraint. Alternatively, fixing the code rate fixes the total number of bins. Then, the sequences are placed optimally within each bin such that the corresponding reconstruction sequences minimize the expected distortion. 

In the anonymous networking setup, let the set of active routes at any given time be referred to as a network session. The key idea is to divide the set of all possible network sessions into bins such that, for each bin, there exists a scheduling strategy that would make the sessions within that bin indistinguishable to an eavesdropper. The level of anonymity required determines the number of bins, and the optimal scheduling strategy plays the role of the reconstruction sequence by minimizing the performance loss across sessions within the bin. 

\subsection{Related Work}\label{sec:related}
Although prevention of traffic analysis is a classical problem, a dominant portion of prior research has centered around Internet applications. In that regard, an important countermeasure was provided by Chaum through the concept of the traffic Mix \cite{Chaum:81ACM}. A Mix node uses re-encryption and packet padding to prevent correlation based on contents or lengths across packets. Further, by batching and reordering packets, the Mix provides anonymity of source-destination pairs. Subsequent improvements in the anonymity provided by the Mix included random delaying (Stop-and-Go Mixes \cite{Kesdogan&etal:98IH}) and introducing dummy packets (ISDN Mixes \cite{Pfitzmann&etal:91CDS}). The concept of Mixes was successfully used in designing remailer and proxy systems \cite{Gulcu&Tsudik:96, Reiter&Rubin:98, Danezis&Dingledine&Mathewson:03} for the Internet. 

Although Mixes provide an ideal solution for many Internet applications, when strict constraints on delay or buffer size are imposed, it was shown \cite{Zhu&etal:04PET} that a Mix no longer provided anonymity to long streams of traffic. An alternative approach, designed primarily for multihop wireless networks is that of deterministic scheduling \cite{Radosavljevic&Hajek:92MILCOM}. In \cite{Radosavljevic&Hajek:92MILCOM}, the authors propose a fixed periodic schedule for the entire network, wherein every node adhered to the schedule by transmitting dummy packets whenever actual data was not present. Although the idea of fixed scheduling can be adapted to handle delay constraints, constant transmission of dummy packets is inefficient and furthermore, the centralized synchronous implementation is impractical for ad hoc wireless networks. 

A key component of our approach is the analytical model for anonymity of routes. In mix networks, anonymity has been measured using the size or entropy of the anonymity set (set of possible source-destination pairs) of an observed packet. In the context of this work, the use of anonymity sets has two disadvantages. First, hiding source-destination pairs alone may not be sufficient, the direction of data flow could also reveal critical information. Second, the measure of anonymity needs to cater to streams of packets rather than a single packet \cite{Zhu&etal:04PET}. Our metric for anonymity is based on the information theoretic notion of equivocation, proposed by Shannon \cite{Shannon:49BSTJ}. Previous applications of equivocation measured the secrecy of transmitted data on point-to-point channels \cite{Wyner:75BSTJ, Csiszar&Korner:78IT}, whereas we use equivocation to measure the secrecy of routes in a network. 

Prevention of traffic analysis can also be viewed as the complementary problem to intrusion detection \cite{Axelsson:00TR}, which is another important area in network security. Some of the techniques we use to design anonymous relaying strategies are motivated by prior work on stepping stone detection \cite{He&Tong:07ITsub}. 

\section{Analytical Model}
The main problem addressed in this paper is to design transmission and relaying strategies that are resilient to traffic analysis and use them to characterize the relationship between achievable network performance and the level of anonymity. We consider a specific category of delay sensitive traffic and measure the network performance using achievable packet relay rates from source to destination. 

\subsection{Notation}
Let $\calG = (\calV,\Emsc)$ be a directed graph, where $\calV$ is the set of nodes in the network and $\Emsc\subset \calV\times \calV$ is the set of directed links. If $(A,B)$ is an element of $\Emsc$, then node $B$ can receive transmissions from node $A$. A sequence of nodes $P=(V_1,\cdots,V_n) \in \calV^*$ is a {\it valid path}\footnote{The notation $\calV^*$ refers to $\bigcup_i \calV^i$.} in $\calG$ if $(V_i,V_{i+1})\in \Emsc,~ \forall i<n$. The set of all possible paths in $\calG$ is denoted by $\calP(\calG)$. 

We assume that during any network observation by the eavesdropper, a subset of nodes communicate using a fixed set of paths. This set of paths $\bS \in 2^{\calP(\calG)}$ is referred to as a network {\it session}. The information that we wish to hide from the eavesdropper is the network session $\bS$. We model $\bS$ as an i.i.d. random variable with a probability mass function $\{p(\bs) : \bs\in 2^{\calP(\calG)}\}$. Therefore,the set of all possible sessions is given by
\[ \Smsc = \{\bs\in 2^{\calP(\calG)} : p(\bs) > 0\}.\]

The prior information $p(\bS)$ on sessions can be obtained using the topology and applications of the particular network, and is also available to the eavesdropper.

\begin{figure}[h]
\centerline{
\begin{psfrags} 
\psfrag{s1}[c]{$S_1$}
\psfrag{s2}[c]{$S_2$}
\psfrag{r}[c]{$B$}
\psfrag{d1}[c]{$D_1$}
\psfrag{d2}[c]{$D_2$}
\scalefig{.350920}\epsfbox{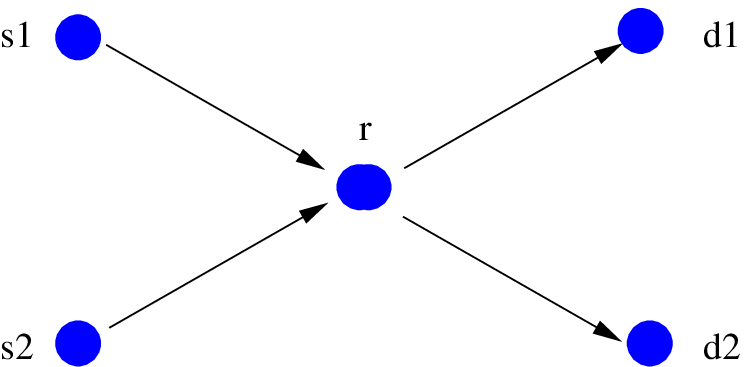} 
\end{psfrags} }
\caption{Two Node Switching Network: $\calG_1 = (\calV,\Emsc)$, $\calV_1 = \{S_1,S_2,B,D_1,D_2\}$, $\Emsc_1=\{(S_1,B),(S_2,B),(B,D_1),(B,D_2)\}$.  \label{fig:two_switch}}
\end{figure}

For example, in a simple network $\calG_1$ as shown in Figure \ref{fig:two_switch}, let $S_1,S_2$ be the only allowed sources and $D_1,D_2$ the allowed destinations. Further, let the sources always communicate with distinct destinations. For such a network, $\calP(\calG_1)$, the set of all possible paths, is given by
\beqq
\calP(\calG_1) =&\{& (S_1,B),(S_1,B,D_1),(S_1,B,D_2),(S_2,B),\\
	      && (S_2,B,D_1),(S_2,B,D_2),(B,D_1),(B,D_2)~\}.
\eeqq
Due to the restriction on distinct destinations, the set of valid sessions $\Smsc$ contains only two sessions:
\beqq
\calS    = &\mbox{\large $\{$}& \left\{(S_1,B,D_1),(S_2,B,D_2)\right\}\\
          &&       \left\{(S_1,B,D_2),(S_2,B,D_1)\right\}\mbox{\large $\}$}.
\eeqq

\noindent{\bf Transmission Schedules} The eavesdropper's observation consists of the
packet transmission epochs in a session. Since it is not possible to determine the location of the eavesdropper(s), we assume that all transmissions are being monitored. Although the packets are encrypted, depending on the physical layer model, it may be possible for an eavesdropper to infer partial information about sender-receiver nodes of packets by merely detecting a transmission. We consider one such physical layer model known as a transmitter directed signaling model.


{\it Transmitter Directed Signaling: }All packets transmitted by a particular node are modulated using the same spreading sequence, and each {transmitting node} is associated with a unique orthogonal spreading sequence. Under this transmission scheme, an eavesdropper would be able to ``tune'' his detector to a particular spreading sequence and detect the transmission times of packets sent by the corresponding node. Although he knows the transmitting node of each packet, we assume that headers are encrypted, so he would not know the intended recipient of any packet. Therefore, in a route involving multiple nodes, even when all transmission schedules are correlated, it is not possible for an eavesdropper to ascertain the final destination node.\\

\noindent{\bf Eavesdropper Observation} Let $\calY_A$ represent the schedule of packets transmitted by node $A$. The schedule $\calY_A$ is a point process,
\[ \calY_A = \{Y_A(1),Y_A(2),\cdots\},\]
where $Y_A(i)$ represents the transmission epoch of the $i^{th}$ packet by node $A$. The eavesdropper detects packet transmission epochs which, by virtue of unique orthogonal codes, would provide him the identity of the transmitting node. Since we assume all nodes are monitored, the eavesdropper's complete observation is given by $\calY = \{\calY_A : A\in \calV\}. $

Note that, while $\calY$ represents the schedules of packet transmissions detected by eavesdroppers, it does not specify which packets are relayed from sources to destinations in a session. In fact, some of the epochs in $\calY$ could represent dummy transmissions by nodes.

\subsection{Anonymity Measure}\label{sec:secrecy}

We model $\calY$ as a random sequence of epochs with conditional distribution $q(\calY|\bS)$. The idea is to design $q(\calY|\bS)$ such that eavesdroppers obtain minimum information about the session $\bS$ by observing $\calY$. Based on the information we wish to hide ($\bS$) and the observation of the eavesdropper ($\calY$), we use equivocation \cite{Shannon:49BSTJ} to define the analytical measure of anonymity.  

\begin{definition}A distribution $q(\calY|\bS)$ is defined to have anonymity $\alpha$ if 
\[ \frac{H(\bS|\calY)}{H(\bS)} \geq \alpha. \]
\end{definition}

\vspace{.6em}

When $\alpha=1$, the distribution $q(\calY|\bS)$ is defined to have {\it perfect anonymity}. For a distribution with perfect anonymity, given the observed schedules, the eavesdropper gains no additional information (than the prior $p(\bS)$) about the routes. In other words,
\[ H(\bS|\calY) = H(\bS).\]

For a general $\alpha$, a physical interpretation of anonymity can be obtained using Fano's Inequality \cite{Cover&Thomas:book}: Let the error probability of the eavesdropper in decoding the session $\bS$ be $P_e$. Then, 
\[ P_e\geq \frac{H(\bS|\calY)-1}{\log|\Smsc|} \geq \frac{\alpha H(\bS)-1}{\log|\Smsc|}.\]

Furthermore, if $\Smsc$ is a large set with uniform prior $\{p(\bs) = \frac{1}{|\Smsc|}, \forall \bs\}$, then
$ P_e \geq \alpha.$ In other words, the anonymity bounds the minimum probability of error incurred by the eavesdropper in decoding $\bS$.

This notion of anonymity that we consider is different from previous definitions \cite{Serjantov&Danezis:02, Kesdogan&etal:98IH}, which were primarily used to hide the source-destination pair of each individual packet. To the best of our knowledge, this is the first definition of anonymity that deals with multihop routes and considers timing information in long streams of transmitted packets.

\subsection{Network Constraints and Throughput}

The key challenge in designing the schedule distribution $q(\calY|\bS)$ with provable anonymity is to sacrifice minimum performance under the networking constraints. In this work, we measure performance using the achievable rates of packets relayed from sources to destinations subject to constraints on medium access and latency, which are described as follows.\\

\noindent{\bf Medium Access Constraints} Wireless networks, due to restrictions on shared bandwidth and transmission power, pose constraints on rates of packets transmitted and received. We consider long streams of packet transmissions, and measure the rate of packets transmitted using an asymptotic measure:
\begin{equation}
 T_A = \lim_{n\rightarrow\infty}\frac{n}{Y_{A}(n)}, \label{equ:tx_rate}
\end{equation}
where $T_A$ denotes the rate of packets transmitted by a node $A$. Since each transmitting node is associated with an orthogonal spreading sequence, the constraint on each point process in $\calY$ is independent.  Specifically, the transmission rate $T_A$ of a node $A$ is bounded by a constant $C_A$, which depends on the characteristics of the medium and the transmission capability of node $A$. As long as $T_A\leq C_A$, successful reception is guaranteed at the intended receiver. 

We assume that the network operates in full duplex mode, where every node can transmit and receive packets simultaneously as long as all transmission rates are within the specified bounds. In other words, a set of schedules $\calY$ is a {\it valid network schedule} if and only if $T_A\leq C_A$ for every node $A$. \\

\noindent{\bf Latency Constraint}: We consider a strict delay constraint on the packets, where the packet delay at each intermediate relay in a route is bounded by $\Delta$. In general, each relay is allowed to reencrypt packets, reorder arrived packets and transmit dummy packets. However, each received data packet at a relay is required to be forwarded within $\Delta$ time units of arrival, or otherwise, dropped. Such a strict delay constraint would apply in practice to time sensitive applications such as target tracking in sensor networks or streaming media in peer to peer networks. In general, a strict delay constraint would prevent congestions in the network and ensure stability, albeit at the cost of dropped packets.

Note that the schedules in $\calY$ only specify when packets are transmitted by each node, and do not indicate which packets actually travel from source to destination on each route of a session. For every schedule, we therefore need to specify a relaying strategy, represented by $\calZ$, which is a set of subsequences of $\calY$. The subsequences  represent the transmissions epochs of packets that are relayed from sources to destinations and therefore, depend on the routes of the session as well as the delay constraint. 

\begin{definition}\label{def:strict}Let a session $\bS = (P_1,\cdots,P_{|\bS|})$, where $P_i$ $=$ $(A(i,1),\cdots,A(i,m(i)))$ is a valid path of length $m(i)$, and $A(i,j)\in \calV$ represents the $j^{th}$ node in path $P_i$ of session $\bS$. A set of subsequences $\calZ = \{\calZ_{i,j} : i\leq |\bS|, j< m(i)\}$ of $\calY$ is a {\it valid relaying strategy} for $\bS$ if:
\ben
\item $\forall i,j$ $\calZ_{i,j} \subseteq \calY_{A(i,j)}$.
\item For every $i,j,n$
\[  0\leq Z_{i,j+1}(n)-Z_{i,j}(n) \leq \Delta. \]
\item If $(A(i,j),A(i,j+1)) = (A(l,m),A(l,m+1))$, then $\calZ_{i,j} \cap \calZ_{l,m} = \phi$.
\een
\end{definition}

\vspace{.77em}

In the above definition, condition 2 ensures that the relayed packets satisfy the delay constraint $\Delta$ at every intermediate relay from the sources to the destinations of the session. Condition 3 ensures that, if any pair of nodes is common to multiple routes, the subsequences picked from the transmission schedules are mutually exclusive. 

In Section \ref{sec:average}, we also consider a relaxed version of the delay constraint, where the average delay of packets is bounded at each relay. The definition for a relaying strategy with average delay constraint can be obtained by modifying condition 2 of Definition 2 as:
\beqa
\forall i,j,n~Z_{i,j+1}(n)-Z_{i,j}(n)&\geq& 0,\label{eq:ave1}\\
\lim_{n\rightarrow\infty}\sum_{m=1}^n \frac{Z_{i,j+1}(m)-Z_{i,j}(m)}{n}&\leq& \bDel.~~\label{eq:ave2}
\eeqa

\begin{figure}[htb]
\centering
\subfigure[$\calY_{X}$ is the transmission schedule of Node $X$.]{ 
 \begin{psfrags} 
 \psfrag{a1}[c]{ $S_1$}
 \psfrag{a2}[c]{ $S_2$}
 \psfrag{b}[c]{ $B$}
 \psfrag{y1}[c]{ $\calY_{S_1}$}
 \psfrag{y2}[c]{ $\calY_{S_2}$}
 \psfrag{yb}[c]{ $\calY_{B}$}
 \scalefig{.273533241730920}\epsfbox{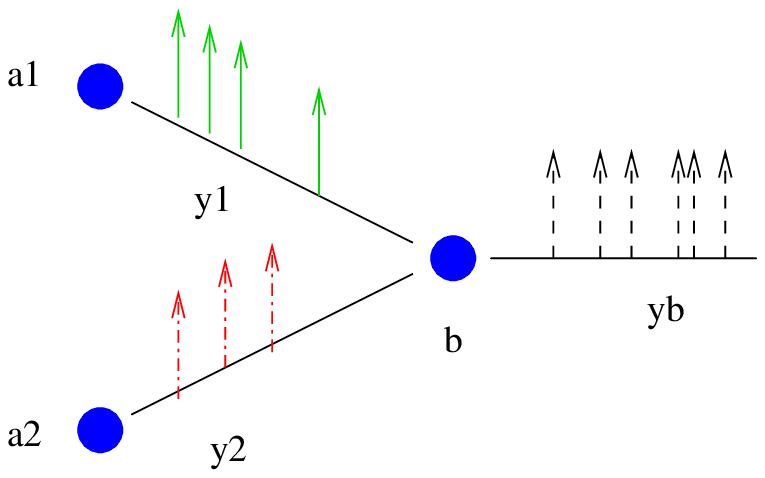} 
 \end{psfrags}} \\
 \vspace{.11em}
 \subfigure[Every packet in the set of subsequences satisfy the strict delay constraint of $\Delta$]{
\begin{psfrags} 
\psfrag{yas}[c]{ $\calZ_{S_1}\cup \calZ_{S_2}$}
\psfrag{ybs}[c]{ $\calZ_B$}
\psfrag{di}[c]{\small $\Delta_i \leq \Delta$}
\scalefig{.352241730920}\epsfbox{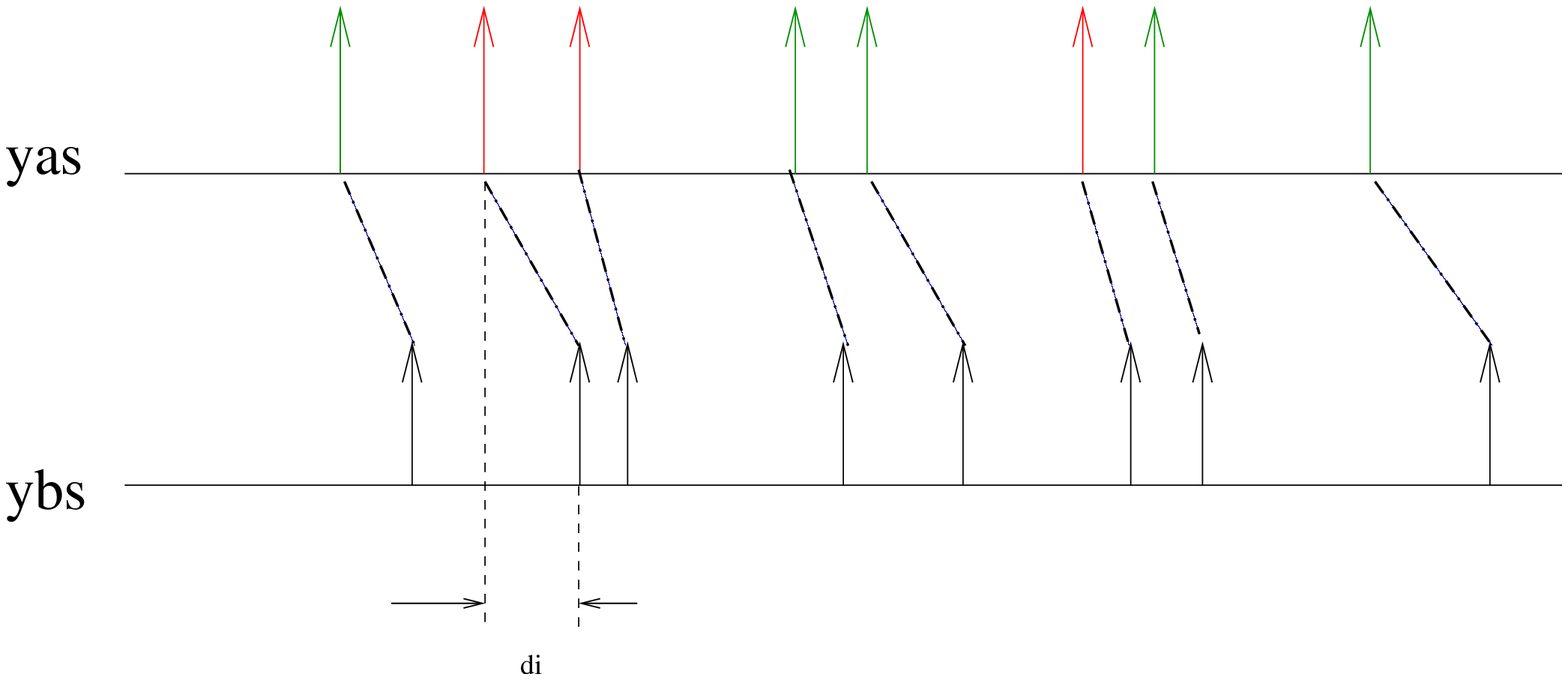} 
 \end{psfrags}} 
\caption{$2\times 1$ Relay with Strict Delay Constraint \label{fig:strict}}
\end{figure}

\subsection{Performance Metrics}\label{sec:performance}

It is possible that the set of subsequences $\calZ$ are a strict subset of the transmissions schedule $\calY$, or in other words, there are epochs in $\calY$ that do not correspond to any relayed packets. Those transmission epochs in $\calY$ that are not present in $\calZ$ would either correspond to packets that are dropped eventually, or represent dummy packet transmissions.  Therefore, for a session $\bs = (P_1,\cdots,P_{|\bs|})$ and relaying schedule $\calZ$, the rate of packets relayed from source to destination on route $P_i$ is given by:
\[ \lm(\calZ,P_i) = \lim_{n\rightarrow\infty}\frac{n}{Z_{i,1}(n)}.\]
Note that, since condition $2$ of Definition 2 ensures that all schedules on a route have same length, it is sufficient to use $\calZ_{i,1}$ to compute rate.

\begin{definition}Let the session vector $\bs = (P_1,\cdots,P_k)$, where $P_i \in V^n$ represents a valid path of data flow. Then, a rate vector $\bflm(\bs) = (\lambda_1,\cdots,\lambda_k)$ is {\it achievable with strict delay} for session $\bs$ if $\exists q(\calY|\bs)$ with anonymity $\alpha$ such that 
\ben
\item Every realization of $\calY$ given $\bs$ is a valid network schedule.
\item For every realization of $\calY$, there exists a valid relaying strategy $\calZ$ that satisfies
\begin{equation}
\lm(\calZ,P_i) \geq \lambda_i,  ~\forall i.
 \end{equation}
\een
\end{definition}

\vspace{.77em}

For a large network with several possible session vectors, characterization of the set of rates for each path of each session vector is potentially cumbersome. Furthermore, in order to draw useful inferences on the relationship between anonymity and network performance, it is helpful to have a simpler quantity representing the achievable performance. We, therefore, propose a scalar metric to characterize the performance of large networks, defined by the average sum-rate as follows. 

\begin{definition}\label{def:sumrate}
$R$ is defined to be a {\it weakly achievable sum-rate with anonymity $\alpha$} if $\exists q(\calY|\bS)$ with anonymity $\alpha$ such that 
\ben
\item For every session $\bs = \{P_1,\cdots,P_{|\bs|}\}$, every realization of $\calY$ given $\bs$ is a valid network schedule.
\item For every realization of $(\bS,\calY)$, there exists a valid relaying strategy $\calZ$, and
\begin{equation}
 \calE\left(\sum_{i=1}^{|\bS|} \lambda(\calZ,P_i)\right) \geq R, \label{equ:ach_rate}
 \end{equation}
where the expectation is over the joint pdf of $\calY$ and $\bS$.
\een
\end{definition}

\vspace{.77em}

Note that the rate and sum-rate defined only represent the rate of packets successfully relayed from sources to destinations. Since the relaying strategy could result in packet drops en route to the destinations, the reliability of the achievable rates needs to be proved by specifying packet encoding and decoding techniques. We address this issue using forward error correction in Section \ref{sec:coding}. 

The fundamental design problem considered in this paper is to characterize the set of achievable rates with anonymity $\alpha$. Specifically, we derive achievability results for two scenarios: For the two hop network (as shown in Fig. \ref{fig:two_hop}), we characterize the set of achievable rate vectors with maximum anonymity ($\alpha=1$) under both delay constraints. For a general network, we use the results from the two hop network and characterize the weakly achievable sum-rate for a general $\alpha$.  

\section{Anonymous Multiaccess Communication}\label{sec:perfect}

\begin{figure}[htb]
\centerline{ 
\begin{psfrags} 
\psfrag{a1}[c]{$S_1$ }
\psfrag{am}[c]{$S_m$ }
\psfrag{d1}[c]{$D_1$ }
\psfrag{dm}[c]{ $D_m$ }
\psfrag{r}[c]{$B$ }
\scalefig{.27520}\epsfbox{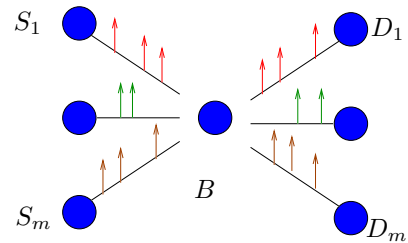} 
\end{psfrags} 
}\caption{Two Hop Network: Source $S_i$ transmits packets to $D_i$ through $B$ \label{fig:two_hop}}
\end{figure}

In this section, we characterize the set of achievable relay rates with maximum anonymity for the two-hop network as shown in Fig. \ref{fig:two_hop}. In particular, we provide rate regions for the session vector $\bs_m = \{(S_1,B,D_1),(S_2,B,D_2),\cdots,(S_m,B,D_m)\}$, $i.e.$ the sources $S_1,\cdots,S_m$ transmit packets to destinations $D_1,\cdots,D_m$ through relay $B$.  

\subsection{Independent Scheduling}\label{sec:pktloss}

In accordance with the definition in Section \ref{sec:secrecy}, scheduling with perfect anonymity corresponds to the independence between session vector $\bS$ and the transmission schedules $\calY$ or in other words,
\[ H(\bS|\calY) = H(\bS) \Rightarrow \bS \perp \calY.\]

We, therefore, propose an independent scheduling technique, wherein each node in the network generates a random transmission schedule, statistically independent of the session and the schedules of other nodes in the network. For example, in the network shown in Fig. \ref{fig:two_hop} with $m=2$
\[ q(\calY|\bS) = q_1(\calY_{S_1})q_2(\calY_{S_2})q_3(\calY_B), \]
where the distributions $q_i$ do not depend on $\bS$. 

Independent scheduling is a particular solution to maintaining anonymity in the two hop setup. An alternative to independent scheduling would be the fixed scheduling as described in \cite{Radosavljevic&Hajek:92MILCOM}. Under that model, all the nodes follow a fixed synchronous schedule irrespective of transmitted data rates or paths of information flow. While the fixed scheduling strategy guarantees maximum anonymity, it would result in a large percentage of dummy packets for low traffic loads. Further, a fixed schedule requires a centralized synchronous implementation, which is impractical in large networks. 

The relaying algorithms discussed in this section are not specific to the statistics of the particular transmission processes and some of the optimal properties hold for any pair of point processes. However, for the purpose of analytical characterization of relay rates, we have modeled the transmission schedules to belong to independent Poisson point processes. Poisson processes have typically been used to model the arrival of packets to nodes in a network, due to memoryless interarrival times property. Although Poisson schedules cannot be shown to be optimal under strict delay constraints, under certain conditions on the physical layer, they are shown to be optimal for an average delay constraint. Our relaying algorithms can be used on other point processes, such as Pareto distributed schedules, however the analytical tractability is not guaranteed. 

\subsection{Scheduling under Strict Delay}\label{sec:strict}

Consider the special case of a single source relay (Fig. \ref{fig:two_hop}, $m=1$). We are interested in the achievable relay rate for the session $\bs_1 = \{(S_1,B,D_1)\}$. The medium access constraints are specified by the bounds $T_{S_1}\leq C_{S_1}, T_B\leq C_B$ on the transmission rates. If the delay constraint was absent ($\Delta=\infty$), then each received packet can be relayed by $B$ at the next available epoch in its transmission schedule. Since packets can be held for an indefinitely long time, the achievable relay rate would be $\lm(\calZ, (S_1,B,D_1)) = \min\{C_{S_1},C_B\}$. Note that this is also the maximum possible rate if node $B$ were to relay packets without any anonymity requirement. 

\begin{figure}[htb]
\centerline{ 
 \begin{psfrags} 
 \psfrag{S1}[l]{Incoming}
 \psfrag{S2}[l]{Outgoing}
 \psfrag{delta}[c]{$\Delta$}
 \psfrag{dr}[c]{Dropped}
 \psfrag{du}[c]{Dummy}
 \scalefig{.357303241730920}\epsfbox{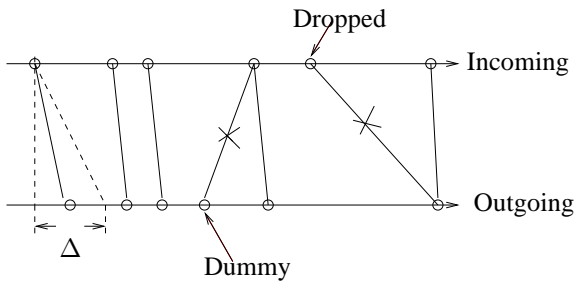} 
 \end{psfrags} 
}\caption{Bounded Greedy Match: Unmatched packets are dropped, unused epochs have dummy packets\label{fig:bgm}} 
\end{figure}

When a strict delay constraint of $\Delta$ is imposed, we design the relaying strategy using the {\it Bounded Greedy Match} (BGM) algorithm proposed in \cite{Blum&Song&Venkataraman:04RAID} under the context of chaff insertion in stepping stone attacks. The algorithm (Fig. \ref{fig:bgm}) is described in Table I. The basic idea is as follows: When a packet arrives at $B$, if there exists a departure epoch within $\Delta$ of the arrival instant and has not been matched to any previous arrival, it is assigned to the arrived packet. Otherwise, the packet is dropped. If a relay epoch does not have any packet assigned to it, the relay transmits a dummy packet at that epoch. 

\begin{table}[h]\label{tab:bgm}
\begin{tabular}{|l|}
\hline
~\\
Let $Y_{S_1}(n),Y_{B}(n)$ represent the arrival time of the $n^{th}$ packet from $S_1$\\
and departure time of $n^{th}$ packet from $B$.\\
\vspace{.3em}
1.~Initialize $i=1,j=1$.\\
2.~Let $t = \min\{Y_{S_1}(i),Y_B(j)\}$.\\
3.~If $t = Y_B(j)$, then \\
~~~~~i. $B$ transmits a dummy packet at time $Y_B(j)$.\\
~~~~~ii. $j=j+1$.\\
~~~else if $Y_B(j)-Y_{S_1}(i)\leq \Delta$\\
~~~~~i. $B$ transmits the $i^{th}$ packet from $S_1$ at $Y_B(j)$.\\
~~~~~ii. $i=i+1,j=j+1$.\\
~~~else \\
~~~~~i. Drop the $i^{th}$ packet that arrived from $S_1$.\\
~~~~~ii.$i=i+1$.\\
4.~Repeat Step 2,3 until the end of the streams.\\
~\\
\hline
\end{tabular}
\caption{Bounded Greedy Match Algorithm}
\end{table}

It was shown in \cite{Blum&Song&Venkataraman:04RAID} that this greedy algorithm resulted in least packet drops. Based on the algorithm, the following theorem characterizes the best achievable relay rate for a pair of independent Poisson processes.

\begin{theorem}\label{thm:single}
If the nodes $S_1$ and $B$ generate independent Poisson transmission schedules, the maximum achievable relay rate from $S_1$ to $D_1$ through $B$ is given by $\lm(\calZ,(S_1,B,D_1)) = C_{S_1} (1-\eps(S_1,B))$ where
\begin{eqnarray}
 \eps(S_1,B) &=& \left\{\begin{array}{cc}  \frac{C_B - C_{S_1}}{C_Be^{-\Delta(C_{S_1}-C_B)}-C_{S_1}} & C_{S_1}\neq C_B\nn\\
\frac{1}{1+C_{S_1}\Delta} & C_{S_1}=C_B\end{array}\right. , \label{equ:single_rtd}\\
            &\defeq& f_e(C_{S_1},C_B).
\end{eqnarray}            

\end{theorem}

\vspace{.6em}

{\it Proof: } Refer to Appendix.

\vspace{.6em}

Theorem \ref{thm:single} expresses the maximum achievable rate in terms of the loss function $\eps(S_1,B)$ where $\eps(S_1,B)$ represents the fraction of packets dropped at relay $B$. As the delay constraint $\Delta$ increases, it is easy to see that the relay rate converges to $\min\{C_{S_1},C_B\}$ which is the optimal rate under no anonymity requirement. Furthermore, the convergence of the relay rate to the optimal value is exponential in $\Delta$. The value of $\eps(S_1,B)$ given in Theorem \ref{thm:single} is obtained when $S_1$ uses the maximum transmission rate of $C_{S_1}$ for this particular route. In a general network, $S_1$ could be simultaneously transmitting to another node, in which case, the rate allocated for $\calY_{S_1,B}$ would be strictly less than $C_{S_1}$. In such a situation, by replacing $C_{S_1}$ in (\ref{equ:single_rtd}) with the allocated rate for the particular flow, we can use Theorem \ref{thm:single} to evaluate the corresponding relay rate.\\

\noindent{\it $m\times 1$ Relay: } For the general $m\times 1$ relay as shown in Fig. \ref{fig:two_hop}, in the absence of the anonymity constraint, the achievable rate region can be obtained using the medium access constraints:
\beq
\bflm(\bs_m) = \{(\lm_1,\cdots,\lm_m) : \lm_i\leq C_{S_i}~ \forall i, \sum_i \lm_i \leq C_B\}.\label{equ:mac} 
\eeq

For a finite delay constraint, a trivial achievable rate region can be obtained if the relay ignores the originating source of the arriving packets. Specifically, the relay uses the BGM algorithm on the joint incoming schedule $\bigcup \calY_{S_i,B}$ and the generated outgoing schedule $\calY_B$. For this strategy, the single source result in Theorem 1 can be easily extended to characterize an achievable rate region for $\bs_m$, which is given in Corollary \ref{cor:hom}. 

\begin{corollary}\label{cor:hom}
There exists a relaying strategy for a $m\times 1$ relay such that the achievable rates $\bflm(\bs_m) = (\lm_1,\cdots,\lm_m)$ satisfy $\lm_i = T_i(1-\eps(S_i,B)), \forall i$ where
\beqa
\eps(S_i,B) &\geq& f_e(\sum_{j=1}^m T_j,C_B), \forall i\\
T_i &\leq& C_{S_i}, \forall i.
\eeqa
\end{corollary}

\vspace{.77em}

\noindent{\bf Prioritized Scheduling}  Ignoring the source identities and considering the joint stream is strictly sub-optimal. Since the relay observes a distinct stream from each source node (by virtue of transmitter directed signaling), the streams can be prioritized to obtain a larger achievable rate region compared to Corollary \ref{cor:hom}. 

Consider a $2\times 1$ relay and assign the highest priority to $S_1$. For every departure epoch in $\calY_{B}$, the relay considers all packets that have arrived within $\Delta$ time units before that epoch. If some of those packets arrived from $S_1$ (highest priority), the relay transmits the earliest of those packets at the chosen epoch. If none of the packets arrived from $S_1$, then the packet that arrived first (from $S_2$) is transmitted. Since $S_1$ is given highest priority, this would provide the maximum rate achievable for the stream from $S_1$. The priority algorithm is formally described in Table II.

\begin{table}[htbp]\label{tab:prior}
\begin{tabular}{|l|}
\hline
~\\
1.~Initialize $i=1,j=1,k=1$.\\
2.~If $Y_B(j)-Z_{S_1,B}(i)\geq\Delta$\\
~~~~~i. Drop $i^{th}$ packet from $S_1$.\\
~~~~~ii. $i=i+1$. Repeat Step $2$.\\
3. Let $t=\min\{Z_{S_1,B}(i),Y_B(j)\}$.\\
4. If $t=Y_{B}(j)$\\
~~~~~i. Let $t' = \min\{Z_{S_2,B}(j),Y_B(k)\}$.\\
~~~~~ii. If $t' = Y_B(k)$ then $B$ transmit dummy packet at $t'$. $k=k+1$.\\
~~~~~~~~~else if $Z_{S_2,B}(j)\geq Y_B(k)-\Delta$ \\
~~~~~~~~~~~~~~$B$ transmits $j^{th}$ packet from $S_2$. $j=j+1, k=k+1$.\\
~~~~~~~~~else \\
~~~~~~~~~~~~~~$j=j+1$. Repeat Step 4.ii.\\
~~~else\\ 
~~~~~~~~~$B$ transmits $i^{th}$ packet from $S_1$. $i=i+1,k=k+1$. \\
5. Repeat Steps 2-4 until end of streams.\\
~\\
\hline
\end{tabular}
\caption{Priority Mapping Algorithm: $S_1$ highest priority}
\end{table}

Similarly, by interchanging the priorities, we can obtain the maximum rate for the stream from $S_2$. It is easy to see that, when none of the sources are given priority, it is equivalent to ignoring the origin of packets (Corollary \ref{cor:hom}). By time-sharing multiple relaying strategies with different priority requirements, a piece-wise linear region of achievable rate vectors is obtained, which is characterized in Theorem \ref{thm:priority}. 

\begin{theorem}\label{thm:priority}
If $\bflm(\bs_2) = (\lm_1,\lm_2)$ represents the achievable relay rates for sources $S_1$ and $S_2$ through relay $B$, then

\noindent 1. $(\lm_1,\lm_2)$ is achievable if
\begin{eqnarray}
\lm_1 \leq a_1 \lm_2 + b_1,~\lm_2 \leq a_2 \lm_1 + b_2,~\lm_i \leq C_{S_i}(1-f_e(C_{S_i},C_B)),
\end{eqnarray}
where $j\neq i$ and
{\scriptsize 
\begin{eqnarray}
a_i =&  \frac{C_{S_i}}{C_{S_j}} + \frac{C_B[(1+\Delta(C_B-C_{S_i}-C_{S_j})-1]}{C_{S_i}(C_Be^{\Delta(C_B-C_{S_i}-C_{S_j})}-C_B)(C_Be^{\Delta(C_B-C_{S_i}-C_{S_j})}-C_{S_i}-C_{S_j})},&~~~~~~~~~~~\\
b_i =& (C_{S_i}-C_{S_j})a_1f_e(C_{S_i}+C_{S_j},C_B).~~~~~~~~~~~~~~~~~~~~~~~~~~~~~~~~~~~~~~~~~~~~~~~~~~~~~~~~&
\end{eqnarray}}

\noindent 2. $(\lm_1,\lm_2)$ is not achievable if
\beq \sum_i \lm_i\geq (C_{S_1}+C_{S_2}) (1-f_e(C_{S_1}+C_{S_2},C_B)),~~\lm_i \geq C_{S_i}(1-f_e(C_{S_i},C_B)).\label{eq:inner}
\eeq

\end{theorem}

\vspace{.6em}

{\it Proof: } Refer to Appendix.

\vspace{.6em}

The priority scheduling cannot be proven to obtain the optimal achievable rate region, and so Theorem \ref{thm:priority} also provides an outer bound to determine the extent of possible sub-optimality. The outer bound is an upper bound on the sum rate $\lm_1+\lm_2$ that is obtained using the optimality of the BGM algorithm. It can be shown that as $\Delta\rightarrow\infty$, the inner and outer bounds coincide and converge exponentially fast. Although the optimality of the region for Poisson processes is still an open problem, the strategy achieves the maximum possible sum-rate. 

The prioritized scheduling can be extended to a general $m\times 1$ relay. Every priority assignment corresponds to an ordering of the sources. When packets from multiple sources contend for a single epoch, the choice of packet to relay is made according to the ordering. Further, by time-sharing strategies for different priority assignments, the complete region can be obtained. 

\begin{figure}[h]
\centerline{ 
\begin{psfrags} 
\psfrag{oa1}[l]{$S_1$ priority $1$}
\psfrag{oa2}[l]{$S_2$ priority $1$}
\psfrag{zer}[l]{priority $0$}
\psfrag{mac}[c]{~~~~No secrecy}
\psfrag{in}[c]{$R_i$}
\psfrag{out}[c]{$R_{out}$}
\scalefig{.36099650920}\epsfbox{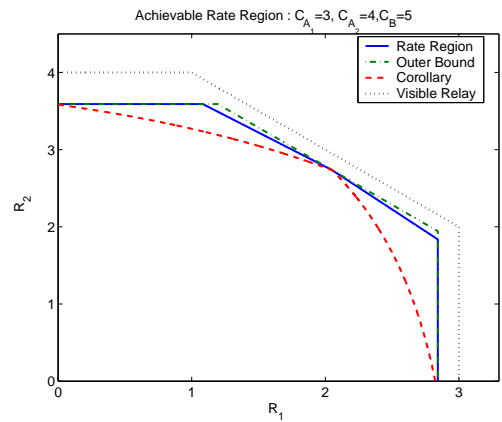} 
\end{psfrags} 
} \caption{$2 \times 1$ Relay rate region. $R_i$ is the rate $\lm(\calZ,(S_i,B,D_i))$. The inner and outer bounds coincide at the maximal sum-rate point.\label{fig:twohop_region}}
\end{figure}

An example region for the $2\times 1$ relay is shown in Fig. \ref{fig:twohop_region}. As is evident, the time-sharing strategy results in a piece-wise linear and convex region. The two corner points of the polygon in the figure which represent the achievable rate-pairs when $S_2, S_1$ are respectively given full priority, clearly demonstrate the gains due to prioritized scheduling. Even when $S_1$ is given full priority, the relay rate for $S_2$ is strictly positive. If no priority is used, however, $S_1$ can achieve maximum rate only when $S_2$ does not transmit at all (region of Corollary \ref{cor:hom}). The maximum priority rate-pairs can also be viewed as the outcome of successive application of the BGM algorithm on the incoming streams from the two sources, with the order of application determined from the priority assignment. 

From theorems $\ref{thm:single}$ and $\ref{thm:priority}$, it is clear that when $C_{S_i},C_B$ and $\Delta$ are finite, the relay rates are strictly less than the transmission rates, thereby resulting in a non-zero packet drop rate. Therefore, the source needs to employ forward error correction (FEC) in order to deliver information to the destination reliably. It can be shown that for very long streams, the coding does not result in further rate reduction (see Section \ref{sec:coding}). 

\subsection{Average Delay}\label{sec:average}
In this section, we consider the average delay constraint at a relay, as specified by (\ref{eq:ave1}) and (\ref{eq:ave2}). It is easy to see that achievable rate regions for an average delay constraint of $\bDel$ can be trivially obtained by using the algorithms of Section \ref{sec:pktloss} that assume a strict delay of $\bDel$. This trivial strategy, however, can be significantly improved by modifying the algorithms appropriately.

Consider the single source relay. Let $m(\Delta, C_{S_1},C_B)$ represent the mean packet delay obtained when the BGM algorithm is applied with strict delay constraint $\Delta$. Since we consider infinitely long streams with an asymptotic constraint, we can choose a strict delay constraint $\Delta^*$ such that the mean delay $m(\Delta^*,C_{S_1},B) = \bDel$. 

\begin{theorem}\label{thm:ave}
$\lm(\calZ,(S_1,B,D_1)) = C_{S_1}(1- \eps(S_1,B))$ is an achievable relay rate for an average delay constraint of $\bDel$ if
\[ \eps(S_1,B) \geq \left\{\begin{array}{cc}  f_e(\Delta^{*},C_{S_1},C_B) & C_B-C_{S_1} \leq \frac{1}{\bDel}\\
					   0          & \mbox{o.w.}\end{array}\right.
					   \]
and $\Delta^*$ is the solution to $m(\Delta^*,C_{S_1},C_B) = \bDel$ where
\[m(\Delta^*,C_{S_1},C_B) = \frac{1+e^{\Delta^*(C_{S_1}-C_B)}\left[\Delta^*(C_{S_1}-C_B)-1\right]}{(C_B-C_{S_1})\left[1-e^{\Delta^*(C_{S_1}-C_B)}\right]}.					 \]
\end{theorem}

\vspace{.4em}

{\it Proof: }Refer to Appendix

\vspace{.6em}

For values of $\bDel$ close to zero, the strict delay constraint $\Delta^* \approx 2\bDel$. Therefore, for very small delays, an average delay constraint does not provide significant improvement in achievable rate compared to a strict delay constraint. However, as $\bDel$ increases beyond a certain threshold, the equivalent strict delay $\Delta^*$  increases exponentially. In that regime, an achievable rate close to optimal can be obtained even for a bounded $\bDel$. Furthermore, as is evident from the Theorem, when $C_B-C_{S_1} \geq \frac{1}{\bDel}$, the strategy achieves zero packet loss. In other words, every transmitted packet can be relayed successfully within the (average) delay constraint. 

\begin{figure}[htb]
\centerline{ 
\begin{psfrags} 
\psfrag{rd1}[c]{$~~\lm^1, \Delta_1^*$}
\psfrag{rd2}[c]{$~~\lm^2,\Delta_2^*$}
\psfrag{rdn}[l]{$\lm^N,\Delta_N^*$}
\scalefig{.3959520}\epsfbox{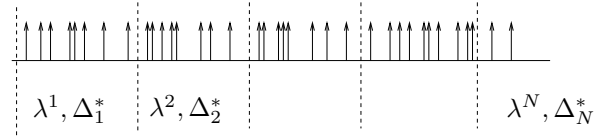} 
\end{psfrags} 
}\caption{Delay Segmentation: In each segment of the traffic, a different strict delay $\Delta_i^*$ is chosen.\label{fig:segment}}
\end{figure}

Since we consider long streams, this strategy could potentially be improved by dividing the stream into finite number ($N$) of segments, and implementing the BGM algorithm with a different strict delay constraint ($\Delta_i^*$) in each segment (see Fig. \ref{fig:segment}). The strict delay constraints should be chosen such that the average delay $\frac{\Sigma_i m(\Delta_i^*,C_{S_1},B))}{N}$ is less than $\bDel$. As the length of the stream increases, each segment $i$ would provide an achievable relay rate $\lm^i=C_{S_i}(1-f_e(\Delta_i^*,C_{S_i},C_B))$  (Theorem \ref{thm:single}) and the net achievable rate would be  $\frac{\Sigma_i \lm^i}{N}$. However, for a pair of Poisson processes, it can be shown that $\lm^i$ is a convex function of the strict delay $\Delta^*_i$, and hence, this segmentation does not reduce\footnote{This convexity may not hold for non-Poisson schedules, in which case, the segmentation could potentially increase the achievable relay rate.} packet loss for a fixed average delay.

Using the relation between the strict delay and average delay in Theorem \ref{thm:ave}, the achievable region for the $m\times 1$ relay can also be obtained by appropriately modifying the strict delay constraint in the prioritized scheduling. The condition on transmission rates for which the priority scheduling strategy is optimal for the $m\times 1$ relay case is a straightforward extension of Theorem \ref{thm:ave}.

\begin{corollary}
There exists a scheduling strategy for average delay $\bDel$ that incurs zero packet loss on all incoming streams,  if the medium access constraints satisfy:
\[ C_B - \sum_i C_{S_i} \geq \frac{1}{\bDel}.\]
\end{corollary}

From the results presented so far, it is clear that while independent Poisson scheduling generally provides a subset of achievable relay rates for strict delay constraints, under certain conditions on the medium access, it can be optimal for an average delay constraint. An important feature in the algorithms presented is that the relays do not require prior knowledge about transmission schedules of the source nodes. The decision to transmit any packet is based on events occurring between its arrival time and the subsequent departure epoch. This makes it particularly attractive for a decentralized implementation of the scheduling, which is of particular value in adhoc wireless and sensor networks. Note that although the rate expressions derived are for Poisson processes, the algorithms presented are quite general, and can be used on any set of point processes. Furthermore, the optimality of the BGM algorithm also holds for any pair of point processes.

\subsection{Reliability}\label{sec:coding}
The independent schedules and relaying algorithms discussed previously result in strictly non-zero packet drop rate for Poisson processes. Further, since the relay nodes generate schedules in a decentralized manner, it is not possible for the source node to know the identities of packets that would be dropped. This implies that the source nodes must employ forward error correction (FEC) techniques to transmit information reliably to the destination. When the traffic is time sensitive such as in media transmission, FEC may not be practical, as it would incur significant coding delay. However, if the strict delay constraint is enforced due to low duty cycles (as in sensor networks) or to maintain stability, it is useful to employ coding to ensure reliability of transmission. 

In order to analyze the reliability of packet transmissions, it is necessary to characterize the channel model between a source and destination. For this purpose, if we treat each packet as a binary unit of data, then the packet drops can be equated to a binary erasure channel. Since packets can be appended with indices, the erasure positions would be known at the destination node. 

Consider a relay node forwarding packets from a single source. Let $E(i)$ denote the random variable indicating that packet $i$ was successfully relayed when applying the BGM relay algorithm. Then, using Proposition 4 in \cite{Boucheron&Salamatian:00IT}, it can be shown that the relay rate obtained from Theorem \ref{thm:single} can be achieved reliably.

\begin{lemma}\label{lemma:coding}
The capacity $C$ of the erasure channel for a single source relay after applying the BGM algorithm is
\[ C = 1- {\lim \sup}_n \frac{1}{n}\sum_{i\leq n} E(i) = 1 - \eps(S_1,B),\]
where $\eps(S_1,B)$ is given by (\ref{equ:single_rtd}).
\end{lemma}
\vspace{.6em}

{\it Proof: } Refer to Appendix.

\vspace{.6em}

The achievability of this reliable rate, however, requires coding across a long stream of packets. Since prioritized scheduling is equivalent to successive application of the BGM algorithm, the rate region of Theorem \ref{thm:priority} also represent reliable rates. In practice, a packet is not a unit of data and the FEC is different from regular point to point communication channels. Coding for packet recovery in networks has been addressed in literature \cite{Shacham&McKenney:90INFOCOM, Rizzo:97ACM}. In particular, in \cite{Shacham&McKenney:90INFOCOM}, the authors propose coding schemes, where, for every block of information packets, parity packets are transmitted such that $\forall i$, the $i$th bit from each packet arranged in sequence forms a codeword from an erasure correcting codebook. 
 
\section{Sum-Rate Secrecy Region}\label{sec:rate_secrecy}
The achievability results presented in the previous section can be viewed as the basic building blocks for hiding routes in a network. While the independent scheduling idea can be directly extended to multihop routes, characterizing rate regions for large networks is cumbersome and not practical. Furthermore, Theorem 2 in \cite{He&Venk&Tong:06MILCOM} shows that under certain conditions, for an $n-$hop path with independent Poisson schedules, the maximum rate of packets that can be relayed to the destination with strict delay constraint decays exponentially as $n$ increases. Therefore, instead of directly extending the idea, we propose to utilize independent scheduling at selected portions of the network depending on the required level of anonymity $\alpha$.

\begin{figure}[htb]
\centerline{ 
\begin{psfrags}  
\psfrag{s1}[c]{  $S_1$ }
\psfrag{s2}[c]{  $S_2$ }
\psfrag{s3}[c]{  $S_3$ }
\psfrag{s4}[c]{  $S_4$ }
\psfrag{d1}[c]{  $D_1$ }
\psfrag{d2}[c]{  $D_2$ }
\psfrag{d3}[c]{  $D_3$ }
\psfrag{d4}[c]{  $D_4$ }
\psfrag{m1}[c]{  $M_1$ }
\psfrag{m2}[c]{  $M_2$ }
\psfrag{m3}[c]{  $M_3$ }
\psfrag{m4}[c]{  $M_4$ }
\scalefig{.359520}\epsfbox{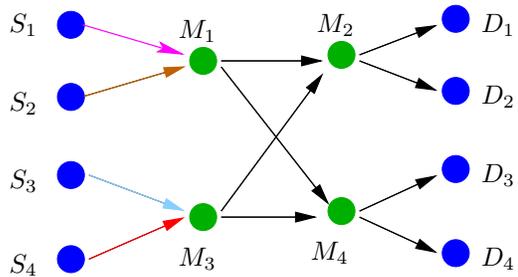} 
\end{psfrags} 
}\caption{Switching Network: Sources $\{S_i\}$ transmit packets to destinations $\{D_i\}$ through relays $\{M_i\}$.\label{fig:switch}}
\end{figure}

As an example, consider the switching network shown in Fig. \ref{fig:switch}. During any network session, each source $S_i$ picks a distinct destination $D_j$. It is easy to see that given the $S_i,D_j$ pairings, there is a unique set of paths in the session $\bS$. If no anonymity is required, each relay would transmit a received packet after a negligible processing delay, thereby incurring no packet drops. Assuming each node has a transmission rate of $C$, the average sum-rate achievable would be $2C$ (min-cut would be out of $M_1,M_3$). Since the schedules of all the relays are dependent on the arrival processes, the eavesdropper would be able to detect the relaying operation of the nodes $M_1,\cdots,M_4$. However, since nodes utilize transmitter directed signaling with encrypted headers, the eavesdropper would not be able to ascertain the final destination nodes of any path. In this case, it can be shown that the anonymity level $\frac{H(\bS|\calY)}{H(\bS)}=.436$.

On the other hand, complete independent scheduling would imply that the relays $M_1,\cdots,M_4$ generate statistically independent schedules. Such a strategy would provide maximum anonymity $\alpha=1$, but result in a reduced achievable sum-rate given by $2C(1-\eps_1)(1-\eps_2)$, where $\eps_1,\eps_2$ are packet losses incurred at relays $M_1,M_3$ and $M_2,M_4$ respectively. 

Suppose, only $M_1,M_3$ were to generate independent schedules, while $M_2,M_4$ relayed packets immediately, the eavesdropper would be able to observe a portion of the paths. In that case, it can be shown that the anonymity level $\frac{H(\bS|\calY)}{H(\bS)} = .65$ (refer to Appendix for details). However since only one relay in each path drops packets, the achievable sum-rate, however, increases to $2C(1-\eps_1)$. 

This simple example illustrates the trade-off between achievable network performance and the level of anonymity. In the remainder of this section, we shall formalize these ideas, describe a randomized relaying strategy and provide an analytical characterization of the achievable sum-rate as a function of anonymity. 

\subsection{Relay Categories}
As suggested in the example, the key idea we exploit is to divide the set of relays according to their scheduling strategies. Specifically, we categorize the relays into two types: {\it covert} and {\it visible} relays.

\begin{figure}[htb]
\centerline{ 
\begin{psfrags} 
\psfrag{vis}[c]{ Visible }
\psfrag{cov}[c]{ Covert }
\psfrag{cor}[c]{ Correlated }
\psfrag{uncor}[c]{ Uncorrelated }
\scalefig{.4359520}\epsfbox{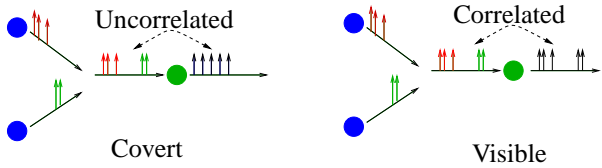} 
\end{psfrags} 
}\caption{Visible and Covert Relaying.\label{fig:relays}}
\end{figure}

{\it Covert Relays: }A relay $M$ is  {\it covert}, if it generates a transmission schedule statistically independent of the schedules of all nodes occurring previously in the paths that contain $M$. For example, if only path $P = \{A_1,\cdots,A_k,M,A_{k+1},\cdots\}$ contains $M$, then $M$ is covert if its transmission schedule is independent of schedules of $A_1,\cdots,A_k$. Further, if $M$ relays packets from $k$ nodes, then it uses the BGM algorithm on the joint incoming packet stream to optimally match the departure epochs. Since our criterion is to maximize sum-rate, the nodes are given equal priority which is the sum-rate optimal strategy (Theorem \ref{thm:priority}).

\begin{figure}[htb]
\centerline{ 
\begin{psfrags} 
\psfrag{vis}[c]{Visible Relay}
\psfrag{cov}[c]{Covert Relay}
\psfrag{y1}[c]{$\calY_1$}
\psfrag{y2}[c]{$\calY_2$}
\psfrag{y3}[c]{$\calY_3$}
\scalefig{.3959520}\epsfbox{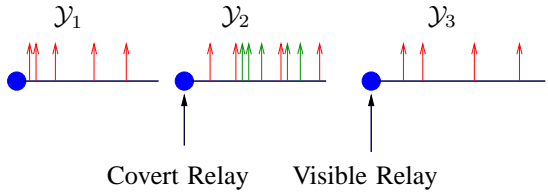} 
\end{psfrags} 
}\caption{Relaying Dummy Packets: $\calY_1$ and $\calY_2$ are statistically independent. If the dummy packets (represented in green) are not relayed, the processes $\calY_1$ and $\calY_3$ will be dependent.\label{fig:visible_dum}}
\end{figure}

{\it Visible Relays: }A {\it visible} relay $M$ generates its schedule based on the schedules of nodes transmitting packets to $M$. For every received packet, the relay schedules an epoch after a processing delay (negligible compared to $\Delta$). It is evident that a relay operating under this highly correlated schedule would be easily detected by an eavesdropper. It is important to note that, although some received packets from the transmitting node may be dummy packets, these are also relayed by a visible node. The reason is that, if dummy packets that were generated due to independent scheduling at a previous node were to be dropped by the visible relay, then the new stream would no longer be independent from the node two hops earlier (see Fig. \ref{fig:visible_dum}). We assume that for visible relays, the eavesdropper makes a perfect detection of the relaying operation. 

By appropriately selecting which relays should be covert in a session, we can guarantee the required level of anonymity. A trivial strategy would be to let all nodes act as covert relays in a session. However, since the independent schedules would result in packet loss at every covert relay, network throughput would be reduced significantly. It is, therefore, necessary to pick the covert relays optimally so that anonymity is guaranteed with minimum loss in throughput. \\

We assume the transmission times of packets by each source node in a session are generated according to an independent Poisson process. To maintain uniformity in traffic schedule patterns, we let the covert relays also generate independent Poisson processes. Given a session $\bS$, let $\bB$ represent the set of relay nodes that are chosen to be covert. Given $\bS,\bB$, using the relaying algorithms discussed in the previous section, the schedules $\calY$ and the relayed subsequences $\calZ$ can be generated for all nodes in the network. 

\subsection{Eavesdropper Observation}

We assume that when a relay is visible, the eavesdropper perfectly correlates the schedules transmitted by a preceding node and the relay. As a result, depending on the set of visible relays, the eavesdropper makes a partial detection on the paths of a session. We denote this partial observation as a set of paths, $\hbS \in 2^{\calP(\calG)}$. Given the observation $\hbS$, the eavesdropper would try and infer the actual session $\bS$.  The partial observation $\hbS$ can be expressed as a function of the actual session $\bS$ and the set of covert relays $\bB$.

We define function $t: 2^{\calP(\calG)}\times \calV \rightarrow 2^{\calP(\calG)}$ to characterize the eavesdropper's observation when at most one relay is covert. For a set of paths $\bP$, $t(\bP,B)$ contains the observed paths when only node $B$ is covert. If $B=\phi$, then $t(\bP,\phi)$ is obtained by removing the destination nodes from every path in $\bP$. This is because, even if all relays are visible, transmitter directed signaling ensures that it is not possible to detect the final destination in any route. If $B\neq \phi$, then a path $P\in \calP(\calG)$ belongs to $t(\bP,B)$ if and only if it satisfies one of the following conditions:

\noindent 1. $\exists P' = (A_1,\cdots,A_k,B,A_{k+1},\cdots,A_{n}) \in \bP$, such that \\
$P = (A_1,\cdots,A_k)$ or $P = (B,A_{k+1},\cdots,A_{n})$.

\noindent 2. $P \in \bP$ and $B\notin P$.

Condition 1 states that, when a path in $\bP$ contains a covert relay, the eavesdropper would observe two different paths, one terminating before $B$ and the other originating from node $B$.  Condition $2$ states that a path that does not contain a covert relay is fully observed. When a subset $\bB = (B_1,\cdots,B_m)\subset V$ of relays are covert, then $\hbS$ can be obtained by repeated application of $t$:
\vspace{-.8em}

\beqa \hbS = t(\cdots(t(t(\bS,\phi),B_1)\cdots),B_m) \defeq \bT(\bS,\bB).\label{equ:eavesd}\eeqa
\vspace{-1.25em}

It can be shown that the set $\hbS$ in the above equation, represents the eavesdropper's sufficient statistic (part of the proof of Theorem \ref{thm:deter}).

\subsection{Throughput Function}

In order to design the optimal selection strategy, we first characterize the loss in sum-rate when a deterministic set of relays are covert in a session. The relaying strategies in Section \ref{sec:pktloss} were designed to minimize the packet loss at a single covert relay. Extending those results to multihop routes, we can characterize the loss in sum-rate of each session $\bS$, when a subset of relays $\bB$ are covert.

If we ignore the anonymity requirement, the best throughput in the network is achieved when all relays are visible. Each session $\bS$ corresponds to a maximum achievable sum-rate obtained using the max-flow that satisfies medium access constraints. Specifically, let ${\bflm}^v(\bS) = (\lm_1^v,\cdots,\lm_{|\bS|}^v)$ represent the vector of achievable relay rates for the paths in session $\bS$ with no covert relays, and $\Lm^v(\bS)$ be the maximum achievable sum-rate. 

If $\bS = (P_1,\cdots,P_{|\bS|})$, then, using the forwarding strategy for visible relays, the maximum achievable sum-rate is the solution to: 
\beqa
\Lm^v(\bS) &=& \max (\lm_1^v+\cdots+\lm_k^v),\label{equ:optim1}\\
 \sum_{i: B\in P_i} \lm_i^v &\leq& C_B,~\forall B\in V.\label{equ:optim3}
\eeqa
 
Therefore our performance metric when anonymity $\alpha=0$ is the maximum expected sum-rate given by, 
\[R(\alpha=0) = \calE(\Lm^v(\bS)),\] 
where the expectation is over the prior $p(\bS)$. Although in practice, the actual rates of flows are dependent on the nature of data and network application, the maximum sum-rate is a metric that represents the fundamental limits of achievable performance. 

When a subset of relays are covert, the achievable sum-rate in each session is reduced depending on the fraction of packets dropped at each covert relay. The net relay rate for each path is obtained by multiplying the fraction of packets that are relayed at every covert relay in that path. 

Specifically, let ${\bf \lm}^c(\bS,\bB) = (\lm_1^c,\cdots,\lm_{|\bS|}^c)$ represent the achievable relay rates from sources to destinations for a session $\bS = (P_1,\cdots,P_{|\bS|})$, when nodes in $\bB$ are covert, and let $\Lm^c(\bS,\bB) \defeq \sum_{i=1}^{|\bS|} \lm_i^c$ be the achievable sum-rate. If $A(i,j)$ represents the $j^{th}$ node in path $P_i$, then
\beqa
\lm_i^c &=& \lm_i^v\prod_{j : A(i,j)\in \bB\cap P_i} \left(1-\eps_i(A(i,j-1),A(i,j))\right).\label{equ:rateloss}
\eeqa
where $\eps_i(A,B)$ represents the fraction of packets transmitted by node $A$ on path $P_i$, that are dropped by covert relay $B$. Note that Theorems \ref{thm:single} and \ref{thm:priority} provide the closed form expression for $\eps_i(A,B)$, if $B$ is the first covert relay in the path $i$. Since the departure epochs of data packets from a covert relay do not constitute a Poisson process, the expression cannot be applied to subsequent covert relays. The analytical characterization of multiple covert relays is generally cumbersome, but can be obtained numerically.

Although the solution of the optimization in ((\ref{equ:optim1}),(\ref{equ:optim3})) specifies a set of transmission rates for the nodes, we know from Theorems \ref{thm:single} and \ref{thm:priority} that, increasing the transmission rates of nodes results in lower packet losses for statistically independent schedules. Therefore, if the relay immediately following a source node is covert, the source node could transmit at the maximum rate possible to minimize packet losses. In other words, if $A$ is a source node, then $T_{A} = \sum_{i:A\in P_i}\lm_i^v$ can be increased to $C_{A}$. Since only the source is allowed to perform forward error correction, it does not help to increase transmission rates of subsequent relays (as we would only get additional dummy packets).  

\section{Performance Characterization}\label{sec:tradeoff}
With the eavesdropper observation of (\ref{equ:eavesd}) and throughput characterization in (\ref{equ:rateloss}), we now have all the elements required to maximize throughput with anonymity $\alpha$. Prior to describing the general randomized strategy, to ease understanding, we first discuss a simple deterministic strategy to obtain a smaller region of achievable  sum-rate anonymity pairs. Then, expanding on that idea, we provide the generalized strategy to characterize the sum-rate anonymity region.

{\it Deterministic Covert Scheduling: } A direct optimization of (\ref{equ:rateloss}) provides a deterministic strategy to characterize achievable sum-rates under anonymity constraints. Specifically, a subset $\bB$ of  relays is chosen to remain covert for all sessions, such that the sum-rate is maximized without violating the anonymity requirement. 

\begin{theorem}\label{thm:deter}
A sum-rate $R$ is achievable with anonymity $\alpha$ if 
\[ R \leq \max_{\bB : H(\bS|\hbS)\geq\alpha} \calE[\blm^c(\bS,\bB)],\]
where $\hbS = \bT(\bS,\bB)$.
\end{theorem}

\vspace{.633em}

{\it Proof: Refer to Appendix}

\vspace{.633em}

Depending on the level of anonymity required, the strategy picks one subset of nodes that are always covert (for all sessions). Since the number of possible subsets is finite, the achievable sum-rate anonymity region would be constant within intervals of $\alpha$, with sudden jumps corresponding to a change in the optimal subset (see example in Section \ref{sec:example}). 

The above theorem provides one set of achievable sum-rates as a function of anonymity $\alpha$. As mentioned in Section \ref{sec:secrecy}, equivocation is an average metric. It gives a lower bound on the {\it average} probability of error for the adversary. Furthermore, the performance is also measured by an average sum-rate metric. Therefore, by time-sharing multiple strategies, it is possible to obtain a convex region without violating the anonymity constraint. 

For example, let two subsets of covert relays $\bB_1$ and $\bB_2$ correspond to achievable sum-rate anonymity pairs $R_1,\alpha_1$ and $R_2,\alpha_2$. At the beginning of every session, one of the subsets $\bB_1,\bB_2$ are chosen with probability $\frac{1}{2}$. Then, it is possible to obtain an achievable sum-rate anonymity pair $(\frac{R_1+R_2}{2}, \frac{\alpha_1+\alpha_2}{2})$. In general, any convex combination of sum-rate anonymity pairs is achievable by time-sharing.

\begin{corollary}
Let 
\[\calR^{\mbox{det}} = \{(R,\alpha) : \mbox{$R$ is an achievable sum-rate with anonymity $\alpha$}\}.\]
Then, every $(R,\alpha)\in \mbox{~convex-hull}(\calR^{\mbox{det}})$ is achievable.
\end{corollary}

\vspace{.6em}

{\it Randomized Covert Scheduling}: The drawback in the strategies discussed above is that the subset $\bB$ is chosen independent of the session $\bS$. The generalized strategy is to chose the set of covert relays as a random function of the session $\bS$. We model the set of covert relays $\bB$ as a random variable with a conditional probability mass function $\{q(\bB|\bS): \bB \in 2^V\}$. The goal is to optimize the conditional p.m.f $\{q(\bB|\bS)\}$ so that achievable sum-rate is maximized for a given level of anonymity $\alpha$. Obtaining the best distribution could typically be done using a brute force optimization over a large dimensional simplex, which is computationally intensive, and impractical for large networks. 
However, the following result proves the duality of this problem to information theoretic rate-distortion, which can then be used to efficiently obtain the optimal strategy and characterize the optimal sum-rate $R(\alpha)$.

\vspace{.4em}

\begin{theorem}\label{thm:ratedist} Let $d: 2^{\calP}\times 2^{\calP} \rightarrow \calR$ s.t.
\begin{equation}
d(\bS,\hat{\bS}) = \left\{\begin{array}{cc} \Lm^v(\bS)-\Lm^c(\bS,\bB) & \exists \bB\mbox{ s.t. } \hat{\bS} = T(\bS,\bB)\\ \infty & \mbox{o.w.}\end{array}\right.~~~~~~~~~~~~~\label{equ:loss_func}
\end{equation}
Then, a sum-rate $R(\alpha)$ is achievable with anonymity $\alpha$ if 
\[ R(0)-R(\alpha) \geq D\left(H(\bS)(1-\alpha)\right), \]
where $D(r)$ is the {\it Distortion-Rate} function defined as
\beq
D(r) = \min_{q(\hat{\bS}|\bS) : I(\bS;\hat{\bS})\leq r} \calE(d(\bS,\hat{\bS})).\label{equ:rate_dist}
\eeq
\end{theorem}
\vspace{.6em}

{\it Proof: } Refer to Appendix.

\vspace{.6em}

The above theorem provides $R(\alpha)$ using the single letter characterization of a rate-distortion function. The loss function $d(\bS,\hbS)$ represents the reduction in sum-rate due to covert relaying. Although the loss function parameters do not explicitly include the set of covert relays $\bB$, it can be shown that given $\bS,\hbS$, the set of covert relays $\bB$ is unique (see proof of Theorem \ref{thm:deter}). Therefore, the distribution $q(\bB|\bS)$ to chose covert relays is equivalent to the distortion minimizing distribution in (\ref{equ:rate_dist}). As a result, the Blahut-Arimoto algorithm \cite{Blahut:72IT} provides an efficient iterative technique to obtain $q(\bB|\bS)$ and the achievable sum-rate $R(\alpha)$. Note that the anonymity $\alpha$ is guaranteed assuming that the eavesdropper is aware of the network topology, the session prior distribution $p(\bS)$ and the optimal strategy $q(\bB|\bS)$ of choosing covert relays. 

\subsection{Discussion}

The equivalence between anonymous networking and rate distortion is not tied to our strategy of choosing covert relays, as explained in Section \ref{sec:results}. In our model, the level of anonymity $\alpha$ directly corresponds to the rate of compression and the performance loss function plays the role of distortion. Therefore, obtaining the optimal rate-distortion function is equivalent to obtaining the throughput anonymity relation.

We believe that the consequences of this duality extend beyond the characterization of the tradeoff between anonymity and throughput. Rate distortion is a field that has been studied for many decades \cite{Cover&Thomas:book}, and the numerous models and techniques developed therein could serve to design strategies for anonymous networking. For example, in our setup, the Blahut-Arimoto algorithm provides an efficient iterative technique to obtain the optimal distribution of covert relays in a session. 

In our current setup, we have considered independent sessions of observation, which may not apply to the scenario where an eavesdropper monitors the network for long periods of time. In that case, we would need a stochastic model to account for session changes, depending on when nodes start or stop communication. Based on the duality we believe that, if we adopt a Markovian model for the session evolution, then techniques in causal source coding \cite{Neuhoff&Gilbert:82IT} would provide possible solutions.

We currently model the entire session as a single entity (the variable $\bS$) which may not be practical to analyze in a large scale network. This model should be broken down to protecting each route independently, depending on the level of anonymity required by that particular route. One approach towards such a model would be to express the set of routes as sequence of links, rather than a single session variable. Each session would then be correspond to a source sequence, and the distortion measure would depend on the relative levels of anonymity required by routes. The challenge in developing such a model, however, is to account for eavesdroppers correlating schedules across multiple hops. 

\section{Example}\label{sec:example}
\begin{figure}[h]
\centerline{ 
\begin{psfrags} 
\scalefig{.5359520}\epsfbox{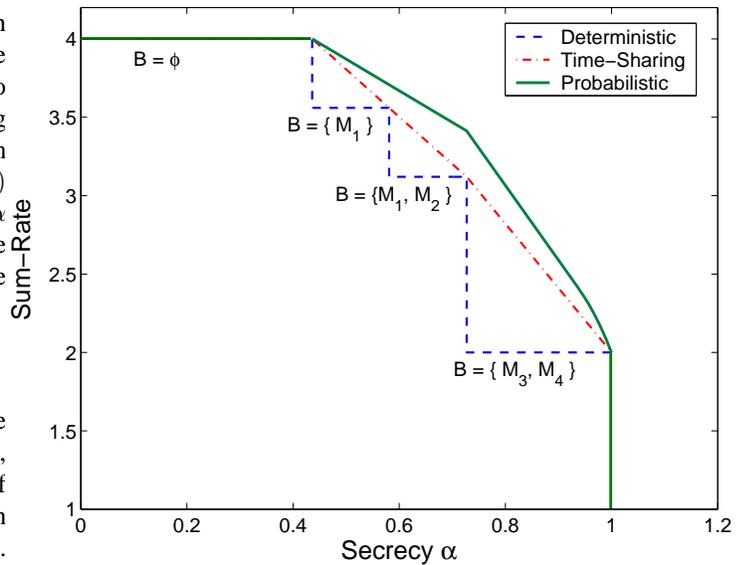} 
\end{psfrags} 
}\caption{Sum-Rate Anonymity Region for $4\times 4$ switching network with $C=2$.\label{fig:rate_secrecy}}
\end{figure}

Consider the switching example given in the beginning of Section \ref{sec:rate_secrecy} (Fig. \ref{fig:switch}). During any network session, each source $S_i$ picks a distinct destination $D_i$. The set of sessions $\calS$, contains 24 elements which are assumed equiprobable. For this example, Fig. \ref{fig:rate_secrecy} plots the sum-rate anonymity region for the deterministic and probabilistic strategies discussed previously. 

The sum-rate anonymity relationship is convex as seen in the figure. This is because the performance metrics, namely anonymity and throughput, are average quantities, which allows time-sharing to convexify any set of achievable rates. The figure clearly demonstrates the performance improvement due to the randomized covert scheduling. As can be seen, when all relays are visible, the maximum sum-rate $2C$ is achieved with a strictly positive secrecy level. This is because, given the transmission stream from relay $M_2$ (or $M_4$), it is not possible for the eavesdropper to detect which packets are received by each destination node. Another interesting observation is that it suffices to make relays $M_2,M_4$ covert in order to obtain perfect anonymity. This shows that, although making all relays covert ensures perfect secrecy, it may not be necessary. 

\section{Conclusions}\label{sec:concl}

One of our key contributions in this work is the theoretical model for anonymity against traffic analysis. To the best of our knowledge, this is the first analytical metric designed to measure the secrecy of {\it routes} in an eavesdropped wireless network. Based on the metric, we designed scheduling and relaying strategies to maximize network performance with a guaranteed level of anonymity. Although we consider specific constraints on delay and bandwidth, the ideas of covert relaying and the randomized selection are quite general, and apply to arbitrary multihop wireless networks. The throughput-anonymity tradeoff we obtain reiterates the known paradigm of inverse relationship between communication rate and secrecy in covert channels.  

In this work, we used throughput as an indicator of network performance and optimized the selection strategy. However, the framework we establish extends beyond maximizing throughput. In fact, the loss function we define in (\ref{equ:loss_func}) can be redefined to represent the loss in any convex function of the achievable relay rates. Further, instead of fixing the packet delay and minimizing the loss in sum-rate, we could fix the rates of transmission and analyze the increase in latency at every covert relay. By optimally designing the loss function to reflect the increase in overall network latency, we would be able to derive  the relationship between latency and level of anonymity. 

\section*{Appendix}

\subsection*{Proof of Theorem \ref{thm:single}}
To prove the theorem, we adopt the technique used in \cite{He&Tong:07ITsub}. Consider the two point processes $\calY_{S_1},\calY_B$. 
Let $X_j$ be the $j$th packet delay, $i.e.$ $X_j=Y_B(j)-Y_{S_1}(j)$.
Define
\[Z_j \: \defeq\: X_j-X_{j-1} =
(Y_{S_1}(j)-Y_B(j-1))-(Y_{S_1}(j)-Y_B(j-1)).\] We see that $Z_j$'s are
i.i.d. random variables; each $Z_j$ is the difference between two
independent exponential random variables with mean $1/C_B$
and $1/C_{S_1}$, respectively. The process
$\{X_j\}_{j=1}^\infty$ is a general random walk with step $Z_j$.
Define $X_0=0$.

Now for every dummy packet transmitted at $t$ in $\calY_B$, we insert
a virtual packet at $t$ in $\calY_{S_1}$; for every packet dropped at time $s$
in $\calY_{S_1}$, we insert a virtual packet at $s+\Delta$ in $\calY_B$. Let the new packet
delays after the insertion of virtual packets be
$\{X'_j\}_{j=0}^\infty$. It can be shown that
$\{X'_j\}_{j=0}^\infty$ is also a random walk with step $Z_j$, but
it has two absorbing barriers at $0$ and $\Delta$, $i.e.$
\[X'_j=\min(\max(X'_{j-1}+Z_j,\: 0),\: \Delta).\]

Since it is almost surely impossible for $X'_{j-1}+Z_j$ to be
exactly equal to $0$ or $\Delta$, each time $X'_j=0$ corresponds
to a dummy transmission in $\calY_B$, and $X'_j=\Delta$ corresponds to a
dropped packet in $\calY_{S_1}$. From example 2.16 in \cite{Cox&Miller:book}, we know that the probability of $X'_j = \Delta$ is given by
\[ \Pr\{X'_j=\Delta\} = \frac{1-\frac{T_{S_1}}{T_B}}{\frac{T_B}{T_{S_1}}e^{-\Delta(T_{S_1}-T_B)}-\frac{T_{S_1}}{T_B}} = \Pr\{X'_j=0\}.\]
Therefore, the fraction of dropped packets in $\calY_{S_1}$ is  
\[ \epsilon_A = \frac{\Pr\{X'_i=\Delta\}}{(1-\Pr\{X'_i=0\})} = \frac{T_B-T_{S_1}}{T_Be^{-\Delta(T_{S_1}-T_B)}-T_{S_1}}.\]

By replacing the transmission rates $T_{S_i},T_B$ with the maximum values $C_{S_i},C_B$, the theorem is proved. In \cite{Blum&Song&Venkataraman:04RAID}, the authors have shown that the BGM algorithm inserts the least chaff fraction for any pair of point processes. Hence, for any $(T_{S_1},T_B)$, it is impossible to obtain a higher information relay rate than (\ref{equ:single_rtd}). This procedure can be extended to multihop by considering multidimensional random walk, but closed form evaluation of the relay rates is cumbersome, even for a few hops.\vspace{-1em}\begin{flushright}$\Box$\end{flushright}
 
\subsection*{Proof of Theorem \ref{thm:priority}}
2. The outer bound is obtained using the optimality of BGM algorithm. Let node $S_i$ transmit at rates $C_{S_i}$. Then, the sum information relay rate obtained by using the BGM algorithm on the joint incoming process is given by:
\beq
\sum_i \lm_i =  (C_{S_1}+C_{S_2})(1-f_e\left(\sum_i C_{S_i},C_B\right)\label{equ:sumrate}.
\eeq

Since BGM inserts the least fraction of dummy packets\cite{Blum&Song&Venkataraman:04RAID}, this is the maximum sum-rate achievable for the given transmission rates. For each individual source $S_i$, the best rate possible is obtained if the other source is completely ignored. Therefore, by replacing $\sum_j C_{S_j}$ by $C_{S_i}$ in (\ref{equ:sumrate}), we can obtain the remaining conditions that specify the outer bound.\vspace{-1.4em} \begin{flushright}$\Box$\end{flushright}

1. Let the zero priority region of Corollary \ref{cor:hom} be represented by $\calR_0$. Every point on the boundary of $\calR_0$, is obtained by letting one node transmit at the highest rate and varying the transmission rate of the other source node from $0$ to the maximum value $C_{S_i}$. This is a special case of priority mapping; the reduced rate for a node is equivalent to marking a fraction of epochs (in a full rate transmission) to be given equal priority. If we forget about the unmarked epochs, then the rate region is identical to Corollary \ref{cor:hom}. However the unmarked epochs owing to unused transmissions in the output schedule still have a chance of being relayed and the BGM algorithm can be used between the unmarked epochs of the input and unused epochs of the output. This successive application of BGM amounts to time-sharing between the zero priority and high priority strategies. Since the point on the boundary of $\calR_0$ has a reduced rate of transmission for one node, it is strictly in the interior of priority achievable rate region. Therefore, the bounding convex polygon forms an inner bound to the best achievable rate region. Evaluating the tangents at the maximum sum-rate point of Corollary \ref{cor:hom} yield the expressions in Theorem 2.\vspace{-1.4em} \begin{flushright}$\Box$\end{flushright}

\subsection*{Proof of Theorem \ref{thm:ave}}

Consider the modified point processes as defined in the proof of Theorem \ref{thm:single}. $X'_i$ denotes the $i^{th}$ step size of the random walk between two absorbing barriers. The average delay incurred by the BGM algorithm is equal to the expected mean size of the random walk without including the steps that hit either boundaries. Following the exposition in example 2.16 in (\cite{Cox&Miller:book}, Page 67), the cumulative distribution of the step size (or delay $\Delta_i$) in the interval $(0,\Delta)$ is given by
\beq
\Pr(X_i\leq x) = \frac{1-\frac{C_{S_1}}{C_B} \exp(\Delta^*+x)(C_{S_1}-C_B)}{1-\frac{C_{S_1}^2}{C_B^2}\exp(\Delta^*(C_{S_1}-C_B))}.\label{equ:stationary}
\eeq
Using the expression above, the average delay $\bDel$ for the BGM algorithm with strict delay $\Delta$ can be evaluated as:
\beqq
\bDel &=& \calE\{X'_i|X'_i\in(0,\Delta^*)\}\\
      &=& \frac{1+\exp(\Delta^*(C_{S_1}-C_B))\left[\Delta^*(C_{S_1}-C_B)-1\right]}{(C_B-C_{S_1})\left[1-\exp(\Delta^*(C_{S_1}-C_B))\right]}.
\eeqq

If $C_B>C_{S_1}$, then as $\Delta^*\rightarrow\infty$,
\beqq
\bDel &=& \frac{1+\exp(\Delta^*(C_{S_1}-C_B))\Delta^*(C_{S_1}-C_B)}{(C_B-C_{S_1})\left[1-\exp(\Delta^*(C_{S_1}-C_B))\right]}\\
      &=& \frac{1}{C_B-C_{S_1}}.
\eeqq
This implies that if $\bDel>\frac{1}{C_B-C_{S_1}}$, then the BGM algorithm with $\Delta^*=\infty$ would be sufficient, and more importantly, optimal. It is easy to see that for small values of $\Delta$, the average delay $\bDel\approx \frac{\Delta^*}{2}$. In other words, when the allowed delay is very small, relaxing the constraint does not provide significant improvement.
\vspace{-1.4em} \begin{flushright}$\Box$\end{flushright}

\subsection*{Proof of Lemma \ref{lemma:coding}}

Consider the modified point processes as defined in the proof of Theorem \ref{thm:single}. $X'_i$ denotes the $i^{th}$ step size of the random walk between two absorbing barriers. Consider a subsequence $\hat{X}_i$ of $X'_i$, wherein $Z'$ contains all points in $X'$ that are strictly greater than $0$. In other words $\hat{X}_i$ does not represent any dummy packets. Accordingly the erasure variable $E(i) = 1_{0<\hat{X}_i<\Delta}$ because a packet is relayed whenever the random walk does not hit either barriers. Since the point processes are renewal processes, the resulting random walk is stationary and the distribution for $X'_i$ given by (\ref{equ:stationary}). Therefore the erasure $E(i)$ is a stationary and ergodic Markov chain and the capacity of the erasure channel is given by
\beqq
 \lim_{n\rightarrow\infty} \frac{1}{n}\sum_i E(i)  &=& 1-{\Pr\{\hat{X}_i=\Delta\}}\\
 						  &=& 1-\frac{\Pr\{X'_i=\Delta\}}{(1-\Pr\{X'_i=0\})}\\
 						  &=& 1-\frac{1-\frac{T_{S_1}}{T_B}}{\frac{T_B}{T_{S_1}}e^{-\Delta(T_{S_1}-T_B)}-\frac{T_{S_1}}{T_B}}\\
 &=& 1 - \eps(S_1,B).
 \eeqq
 
\vspace{-1.4em} \begin{flushright}$\Box$\end{flushright}

\subsection*{Proof of Theorem \ref{thm:deter}}

From (\ref{equ:rateloss}), we know that $\bflm^c(\bS,\bB)$ is an achievable relay rate vector when nodes in $\bB$ are covert. It remains to be seen that the condition $H(\bS|\hbS)\geq \alpha$ guarantees an anonymity $\alpha$. For this purpose, it is sufficient to show that
\[ H(\bS|\calY) \leq H(\bS|\hbS).\]

Let $\hcY$ be the schedules generated assuming $\hbS$ was a session and none of the nodes were covert. The transmission rates of nodes in $\hcY$ are assumed identical to $\calY$. For the nodes that are the sources in $\bS$, the schedules are independent in $\calY$ and $\hcY$. Session $\hbS$ has additional sources due to the broken paths, which also generate independent transmission schedules. The set of these additional sources is identical to the set of covert relays in $\bS$. Therefore, the schedules are independent in $\calY$ as well. Since the remaining nodes relay all received packets within negligible processing delay, $p(\calY|\bS) = p(\hcY|\bS)$. Then, using the data processing inequality ($\bS-\hbS-\hcY$)
\[
H(\bS|\calY) = H(\bS|\hcY) \leq H(\bS|\hbS).\]

\vspace{-1.4em} \begin{flushright}$\Box$\end{flushright}

\subsection*{Proof of Theorem \ref{thm:ratedist}}

Consider the optimal solution $q^*(\hbS|\bS)$ of the distortion rate problem,
\[ D = \min_{q(\hat{\bS}|\bS) : I(\bS;\hat{\bS})\leq (1-\alpha)H(\bS)} \calE(d(\bS,\hat{\bS})).\]

From the definition of $d(\bS,\hbS)$, it is easy to see that if $\nexists \bB ~s.t.~ \hbS = \bT(\bS,\bB)$, then $q^*(\hbS|\bS)=0$. Given $\bS,\hbS$, we can show that the set of covert relays $\bB$ are uniquely determined, using the following argument: 

Suppose $\exists \bB_1\neq\bB_2$ such that $\bT(\bS,\bB_1)=\bT(\bS,\bB_2)$. Then, we can write $\bB_1 = (\bB,\bB_1'),\bB_2=(\bB,\bB_2')$ where $\bB_1' = (B_{11},\cdots,B_{1m})$, $\bB_2' = (B_{21},\cdots,B_{2n})$ and $\bB_1'\bigcap\bB_2' = \phi$. We know that
\beqq 
\hbS(\bS,\bB_1) &=& t(\cdots t(\bT(\bS,\bB),B_{11}),\cdots),B_{1m})\\
		&=& t(\cdots t(\bT(\bS,\bB),B_{21}),\cdots),B_{2n}) = \hbS(\bS,\bB_2).
\eeqq

Suppose none of the paths in $\bT(\bS,\bB)$ contain $\bB_1'\bigcup\bB_2'$, then it does not matter if those relays are covert or not, in which case the subset of covert relays would be $\bB$.

If $\exists P \in \bT(\bS,\bB)$ that contains  $B_{11}$, then $\bT(\bS,\bB_1)$ would contain a path that ends in $B_{11}$, whereas $\bT(\bS,\bB_2)$ cannot contain such a path. Therefore, we have a contradiction.

The above argument shows that we can equivalently write $q^*(\hbS|\bS) = q^*(\bB|\bS)$. Therefore, $q^*$ specifies a valid selection strategy.  Since $H(\bS)$ is fixed apriori, $I(\bS;\hbS)\leq (1-\alpha)H(\bS)$ ensures that an anonymity $\alpha$ is guaranteed. Further, for every $\bB$, the function $d$ evaluates the difference in achievable rate vectors $\bflm^v(\bS)$ and $\bflm^c(\bS,B)$. Taking expectation over $q^*(\bB|\bS)$, it is easy to see that the distortion $D$ is achievable with $\alpha-$anonymity. \vspace{-1.1em}\begin{flushright}$\Box$\end{flushright}

\noindent{\it Switching Network Example}

When all relays are visible, the eavesdropper would not know the final node of any route. This implies that given an observation, $4$ possible source-destination pairings would be equally likely. This implies that his uncertainty $H(\bS|\calY) = \log(4)$. Since the priors are equally likely $H(\bS) = \log(24)$. Therefore, when all relays are visible, $\alpha = \frac{\log(4)}{\log(24)} = .436$.\\

\noindent When $M_1,M_3$ are covert, the number of possible pairings given an observation would depend on the session. For example, if $\{(S_1,M_1,M_2,D_1)$,$(S_2,M_1,M_2,D_2)$,$(S_3,M_3,M_4,D_3)$,\\
$(S_4,M_3,M_4,D_4)\}$ is the session, then the eavesdropper would be able to identify that all transmissions from $M_1$ are relayed by $M_2$, and his uncertainty would be $\log(4)$. This is identical to $7$ other pairings (whenever $S_1,S_2$ use the same set of relays). Suppose $\{(S_1,M_1,M_2,D_1),(S_2,M_1,M_4,D_3)$,$(S_3,M_2,M_3,D_2)$,\\
$(S_4,M_2,M_4,D_4)\}$ was the session, then it would be indistinguishable from the $15$ remaining sessions (whenever $S_1,S_2$ do not use the same set of relays), and his uncertainty would increase to $\log(16)$. Therefore, since all sessions are equally probable,
\[\frac{H(\bS|\calY)}{H(\bS)} = \frac{(1/3)\log(4)+(2/3)\log(16)}{\log(24)} = 0.659.\]

{\small 
\bibliographystyle{ieeetr}
\bibliography{required_files/References}}

\begin{thebibliography}{10}

\bibitem{West:Book}
N.~West, {\em The SIGINT Secrets: The Signal Intelligence War: 1900 to Today}.
\newblock New York: William Morrow, 1988.

\bibitem{Voydock&Kent:83ACM}
V.~L. Voydock and S.~T. Kent, ``Security mechanisms in high-level network
  protocols,'' {\em ACM Computing Surveys}, vol.~15, pp.~135--171, 1983.

\bibitem{Raymond:01}
J.-F. Raymond, ``Traffic analysis: Protocols, attacks, design issues and open
  problems,'' in {\em Designing Privacy Enhancing Technologies: Proceedings of
  International Workshop on Design Issues in Anonymity and Unobservability}
  (H.~Federrath, ed.), vol.~2009 of {\em LNCS}, pp.~10--29, Springer-Verlag,
  2001.

\bibitem{Sun&etal:02}
Q.~Sun, D.~R. Simon, Y.~Wang, W.~Russell, V.~N. Padmanabhan, and L.~Qiu,
  ``Statistical identification of encrypted web browsing traffic,'' in {\em
  Proceedings of the 2002 IEEE Symposium on Security and Privacy}, (Berkeley,
  California), p.~19, May 2002.

\bibitem{Mathewson&Dingledine:04PET}
N.~Matthewson and R.~Dingledine, ``Practical traffic analysis: Extending and
  resisting statistical disclosure,'' in {\em Privacy Enhancing Technologies:
  4th International Workshop}, May 2004.

\bibitem{Felten&Schneider:00CCS}
E.~W. Felten and M.~A. Schneider, ``Timing attacks on web privacy,'' in {\em
  {ACM} Conference on Computer and Communications Security}, pp.~25--32, 2000.

\bibitem{Song&Wagner&Tian:01}
D.~X. Song, D.~Wagner, and X.~Tian, ``{Timing Analysis of Keystrokes and Timing
  Attacks on SSH},'' in {\em Proc. 10th USENIX Security Symposium},
  (Washington, DC), August 2001.

\bibitem{Shannon:49BSTJ}
C.~E. Shannon, ``Communication theory of secrecy systems,'' {\em Bell System
  Technical Journal}, 1949.

\bibitem{Chaum:81ACM}
D.~Chaum, ``{Untraceable electronic mail, return addresses and digital
  pseudonyms},'' {\em Communications of the ACM}, vol.~24, pp.~84--88, February
  1981.

\bibitem{Kesdogan&etal:98IH}
D.~Kesdogan, J.~Egner, and R.~Buschkes, ``{Stop-and-go MIXes providing
  probabilistic security in an open system},'' in {\em Second International
  Workshop on Information Hiding (IH'98), Lecture Notes in Computer Science},
  vol.~1525, (Portland, Oregon), pp.~83--98, April 1998.

\bibitem{Pfitzmann&etal:91CDS}
A.~Pfitzmann, B.~Pfitzmann, and M.~Waidner, ``{ISDN-MIXes: Untraceable
  communication with very small bandwidth overhead},'' in {\em Proceedings of
  the GI/ITG Conference: Communication in Distributed Systems,
  Informatik-Fachberichte}, vol.~267, (Mannheim, Germany), pp.~451--463,
  February 1991.

\bibitem{Gulcu&Tsudik:96}
C.~Gulcu and G.~Tsudik, ``Mixing e-mail with babel,'' in {\em Proceedings of
  the Symposium on Network and Distributed System Security}, pp.~2--19,
  February 1996.

\bibitem{Reiter&Rubin:98}
M.~K. Reiter and A.~D. Rubin, ``Crowds: anonymity for {Web} transactions,''
  {\em ACM Transactions on Information and System Security}, vol.~1, no.~1,
  pp.~66--92, 1998.

\bibitem{Danezis&Dingledine&Mathewson:03}
G.~Danezis, R.~Dingledine, and N.~Mathewson, ``Mixminion: design of a type iii
  anonymous remailer protocol,'' in {\em Proceedings of 2003 Symposium on
  Security and Privacy}, pp.~2--15, May 2003.

\bibitem{Zhu&etal:04PET}
Y.~Zhu, X.~Fu, B.~Graham, R.Bettati, and W.~Zhao, ``On flow correlation attacks
  and countermeasures in mix networks,'' in {\em Proceedings of Privacy
  Enhancing Technologies workshop}, May 26-28 2004.

\bibitem{Radosavljevic&Hajek:92MILCOM}
B.Radosavljevic and B.~Hajek, ``Hiding traffic flow in communication
  networks,'' in {\em Military Communications Conference}, 1992.

\bibitem{Wyner:75BSTJ}
A.~Wyner, ``{The wiretap channel},'' {\em Bell Syst. Tech. J.}, vol.~54,
  pp.~1355--1387, 1975.

\bibitem{Csiszar&Korner:78IT}
I.~Csisz\'{a}r and J.~Korner, ``Broadcast channels with confidential
  messages,'' {\em IEEE Trans. on Information Theory}, vol.~24, pp.~339--348,
  May 1978.

\bibitem{Axelsson:00TR}
S.~Axelsson, ``{Intrusion detection systems: A taxonomy and survey},'' tech.
  rep., Chalmers University of Technology, Sweden, March 2000.

\bibitem{He&Tong:07ITsub}
T.~He and L.~Tong, ``{Detecting Information Flows: Fundamental Limits and
  Optimal Algorithms}.'' submitted to IEEE Trans. on Information Theory, 2007.

\bibitem{Cover&Thomas:book}
T.~Cover and J.~Thomas, {\em Elements of Information Theory}.
\newblock John Wiley \& Sons, Inc., 1991.

\bibitem{Serjantov&Danezis:02}
A.~Serjantov and G.~Danezis, ``Towards an information theoretic metric for
  anonymity,'' in {\em Proceedings of Privacy Enhancing Technologies Workshop
  (PET 2002)} (R.~Dingledine and P.~Syverson, eds.), Springer-Verlag, LNCS
  2482, April 2002.

\bibitem{Blum&Song&Venkataraman:04RAID}
A.~Blum, D.~Song, and S.~Venkataraman, ``{Detection of Interactive Stepping
  Stones: Algorithms and Confidence Bounds},'' in {\em Conference of Recent
  Advance in Intrusion Detection (RAID)}, (Sophia Antipolis, French Riviera,
  France), September 2004.

\bibitem{Boucheron&Salamatian:00IT}
S.~Boucheron and M.~R. Salamatian, ``{About Priority Encoding Transmission},''
  {\em IEEE Transactions on Information Theory}, vol.~46, pp.~699--705, March
  2000.

\bibitem{Shacham&McKenney:90INFOCOM}
N.~Shacham and P.~McKenney, ``{Pakcet Recovery in High-Speed Networks using
  Coding and Buffer Management},'' in {\em Proc. IEEE INFOCOM}, pp.~124--131,
  1990.

\bibitem{Rizzo:97ACM}
L.~Rizzo, ``{Effective Erasure Codes for Reliable Computer Communication
  Protocols},'' in {\em Proc. ACM SIGCOMM Computer Communication Review},
  vol.~27, pp.~24--36, 1997.

\bibitem{He&Venk&Tong:06MILCOM}
T.~He, P.~Venkitasubramaniam, and L.~Tong, ``{Packet scheduling against
  stepping-stone attacks with chaff},'' in {\em Proc. IEEE Military
  Communications Conference}, (Washington,DC), October 2006.

\bibitem{Blahut:72IT}
R.~Blahut, ``{Computation of Channel Capacity and Rate-Distortion Functions},''
  {\em IEEE Trans. Infor. Theory}, vol.~IT-18, July 1972.

\bibitem{Neuhoff&Gilbert:82IT}
D.~Neuhoff and L.~Gilbert, ``{Causal Source Codes},'' {\em IEEE Trans. on
  Information Theory}, vol.~28, pp.~701--713, Sep. 1982.

\bibitem{Cox&Miller:book}
D.~Cox and H.~Miller, {\em The Theory of Stochastic Processes}.
\newblock New York: John Wiley $\&$ Sons Inc., 1965.

\end{thebibliography}

\end{document}